\documentclass[preprint, 3p, 12pt]{elsarticle}
\usepackage[utf8]{inputenc}  % UTF-8 input encoding
\usepackage[T1]{fontenc}     % Type1 fonts
\usepackage{lmodern}         % Improved Computer Modern font
\usepackage{microtype}       % Better handling of typo
\usepackage[scaled]{helvet} % Scale Helvetica
\usepackage[english]{babel}  % Hyphenation
\usepackage{graphicx}        % Import graphics
\usepackage[bookmarks=false]{hyperref}            % Better handling of references in PDFs
\hypersetup{pdfauthor={Name}}
\usepackage{comment}         % To comment out blocks of text
\biboptions{sort&compress}   % Sort and compress citations

\journal{Elsevier}

\usepackage{amsmath,amsfonts,amssymb}

\usepackage{subfig}

\usepackage{xcolor}
\usepackage{todonotes}
\usepackage{graphicx}
\usepackage{epstopdf, epsfig}
\usepackage{mathtools}
\usepackage{lipsum}
\usepackage{array}
\usepackage{accents}
\usepackage{amsmath}
\usepackage{xparse}
\usepackage{nicematrix}
\usepackage{stackengine}

\usepackage{xcolor}
\usepackage{tikz}
\usetikzlibrary{fit}
\DeclareMathOperator*{\argmin}{argmin}
\usepackage{wrapfig}
\usepackage{multirow}
\usepackage{makecell}
\usepackage{tabularx}
\usepackage{lscape}
\usepackage{dcolumn}
\newcolumntype{d}{D{.}{.}{2.5}}
\usepackage{siunitx}
\usepackage{bm}
\usepackage{soul}
\usepackage{placeins}
\usepackage{scalerel}
\newcommand{\Tau}{\scalerel*{\tau}{X}}
\usepackage{rotating}

\NewDocumentCommand{\underarrow}{O{=} O{\downarrow} m}{%
  \underset{\makebox[0pt]{\begin{tabular}{@{}c@{}}\ensuremath{#2}\\[0pt]#3\end{tabular}}}{#1}
}
\graphicspath{{fig/}}

\begin{document}

\begin{frontmatter}

% Title

\title{Discovery of Sparse Invariant Subgrid-Scale Closures via Dissipation-Controlled Training for Large Eddy Simulation on Anisotropic Grids}

% Authors and affiliations

\author{Samantha Friess\corref{cor1}\fnref{fn1}}
\ead{samantha.friess@colorado.edu}
\cortext[cor1]{Corresponding author}

\author{Aviral Prakash\fnref{fn2}}

\author{John A. Evans\fnref{fn1}}

\fntext[fn1]{University of Colorado Boulder, Boulder, CO 80309, USA}
\fntext[fn2]{Los Alamos National Laboratory, NM 87545, USA}
    
\begin{abstract}
%% ----------------------
%% Write your abstract here. Do not enclose it in an "abstract"
%% environment.
%% ----------------------

Neural networks offer highly expressive turbulence closures, yet their complexity obscures the physical mechanisms they aim to model, and their computational cost can limit their tractability. To address these limitations, we introduce a sparsity-promoting subgrid-scale (SGS) stress closure modeling framework that identifies explicit polynomial model forms using sparse regression. Candidate models are constructed through scaling a minimal tensor basis by a truncated polynomial expansion of invariant scalars, thereby enforcing fundamental invariance properties while regulating the highest order of admissible terms. Arbitrary filter anisotropy is incorporated to enable consistent representation of turbulent structures across computational grids with anisotropic scales and resolutions. We also explicitly constrain SGS energy dissipation during training to improve functional performance and promote numerical stability. The framework is trained on a small dataset of idealized turbulence and evaluated through a series of \textit{a priori} and \textit{a posteriori} tests. Sensitivity studies examine the effects of variations in model order and optimization penalties for regularization and dissipation across a range of canonical flow configurations beyond those represented in the training dataset. We also evaluate on a separated flow benchmark to assess generalizability to a more complex turbulent regime. In many cases, the sparse regression closures achieve predictive accuracy comparable to an invariance-preserving neural network while retaining markedly simpler parametric forms. Moreover, we demonstrate that the sparse closures can be trained and evaluated at a fraction of the cost of the neural network model.

%% :folding=explicit:wrap=soft:mode=latex:

\end{abstract}
\end{frontmatter}

%% ---------------------------
%% If you wish to include additional packages, define new environments or
%% new commands, put them in the file includes.tex
%%
%% Write your abstract in the file abstract.tex.
%% ---------------------------

%% ---------------------------
%% Introduction
%% ---------------------------
\section{Introduction}\label{sec:intro}

Machine learning and data-driven techniques have gained increasing attention within the turbulence closure modeling and simulation community in recent years \citep{Duraisamy19, Brunton20, Beck21, Zhang23, Sanderse25}. These approaches offer the potential for improved accuracy and generalization across different flow cases and features by leveraging large datasets generated from high-fidelity simulations or experimental measurements. In practical engineering applications, the ability of a turbulence closure model to generalize to new flow conditions is crucial \citep{Brunton20, Taghizadeh20, Beck21, Kutz22, Girimaji24, Sanderse25}. A key element shared among the most robust, generalizable, and accurate data-driven models outlined in the literature is the direct integration of physical symmetry and invariance properties into their model forms \citep{Duraisamy19, Brunton20, Sanderse25}. Incorporating invariance properties, such as Galilean, unit, reflectional, and rotational invariance, ensures that closure models adhere to the governing principles of turbulent fluid flow \citep{Ling16}. This facilitates a more accurate depiction of the turbulent physics during both interpolation and extrapolation, thereby improving the models' predictive performance. 

A large body of data-driven turbulence closures with embedded invariance properties use neural networks as the machine learning model. Neural networks are an attractive option given their ability to effectively represent highly nonlinear dynamics, which has enabled the data-driven turbulence models that utilize them to consistently outperform classical models \citep{Ling16, Parmar20, Xie20, Strofer21, Peters22, Prakash22, Prakash24, Wu24}. Nonetheless, neural networks have several notable limitations: a lack of interpretability owing to their "black box" nature that obstructs the model's underlying physical principles, a susceptibility to improperly fitting the model if not fed adequately extensive quantities of high-fidelity data, and a substantial computational expense during the training and evaluation routines that may be prohibitive. In response, alternative machine learning approaches --- including sparsity-promoting and symbolic regression methods such as evolutionary algorithms, relevance vector machines, and sparse regression --- have been explored to improve model transparency, generalizability, and tractability. By enforcing parsimony, these methods identify a compact subset of the most relevant variables that best capture the underlying physics, yielding explicit algebraic expressions whose contributions can be traced to facilitate physical insight. However, given the inherent trade-off between model complexity, efficiency, and accuracy, sparsity-promoting closure models proposed in the literature exhibit a wide range of applicability, reliability, and stability characteristics.

In the Reynolds-averaged Navier–Stokes (RANS) setting, numerous sparsity-promoting methodologies have been explored to formulate Reynolds stress closures. Most of these approaches focus on learning targeted corrections to established RANS closures, seeking to improve predictive performance in deficient regions while aiming to preserve the robustness of the baseline model. Note, however, that the numerical stability of the baseline model is not guaranteed in such approaches \cite{Weinmann09, Sanderse25}. Early work by \cite{Weatheritt16, Weatheritt17} used Gene Expression Programming (GEP) --- an evolutionary symbolic regression technique --- to augment a baseline linear eddy viscosity (LEV) model with the addition of an extra learned anisotropy tensor correction term. Subsequent work by \cite{Chakrabarty21} used this same approach to further demonstrate the suitability of GEP for enhancing the performance of existing turbulence closures. Building on this GEP methodology, \cite{Zhao20} introduced the so-called CFD-driven training approach, where the fitness of candidate GEP-based closure corrections is evaluated by integrating a RANS solver into the training process. CFD-driven symbolic regression using GEP has also been explored for Reynolds stress augmentation in turbulent mixing problems \citep{Xie22}, for turbulent heat flux enhancement in gas turbine trailing-edge slot flows \citep{Lav21}, and for multi-objective optimization of concurrently trained model corrections \citep{Waschkowski22}. In related work, \cite{Liao25} used data assimilation in the training phase to mutually couple a RANS solver with a stress correction model developed via genetic algorithms. Alternative sparsity-promoting machine learning methodologies have been applied in the RANS context. Notably, \cite{Schmelzer20}'s Sparse Regression of Turbulent Stress Anisotropy (SpaRTA) framework, which enables corrections to both the anisotropy tensor and turbulent transport equations of baseline RANS closures. Extensions of SpaRTA include applications to wind turbine wake modeling \citep{Steiner22}, formulations incorporating Sparse Bayesian Learning (SBL) for probabilistic sparsity and uncertainty quantification \citep{Cherroud22}, and CFD-driven coupling to embed a CFD solver within the training process \citep{BenHassanSaidi22}. Furthermore, recent hybrid schemes have combined deep learning with symbolic regression to learn data-driven enhancements to pre-existing RANS closures: \cite{Tang23a} employed the deep symbolic regression (DSR) \cite{Petersen21} method for model discovery and \cite{Alhafiz25} used Kolmogorov–Arnold Networks (KANs) \cite{Liu25}. Other recent efforts have employed field inversion of a baseline turbulence model to derive symbolic correction factors for improved separated flow predictions \cite{Wu23, Wu25}. Although many of these correction-oriented RANS closure formulations have demonstrated improvements in \textit{a posteriori} tests over their baseline turbulence models, their reliance on a pre-existing closure imposes inherent limitations due to the empirical assumptions embedded in those models \cite{Taghizadeh20, Cherroud22, Mandler24}. To overcome these limitations, \cite{Beetham20} employed sparse regression to construct stand-alone algebraic representations of the Reynolds stress tensor, later extending their approach to multiphase flows \cite{Beetham21}.

Sparsity-promoting and symbolic regression techniques have also been explored in the large eddy simulation (LES) setting to develop stand-alone, explicit subgrid-scale (SGS) closures. GEP-based machine learning models were utilized in early data-driven LES closure studies by \cite{Li21} and \cite{Reissmann21} to derive explicit algebraic SGS models for canonical turbulence cases, with the former performing offline training, and the latter incorporating a CFD solver into the optimization loop for iterative model coefficient refinement. An extension of the Sparse Identification of Nonlinear Dynamics (SINDy) framework --- a sparse regression-based governing equation discovery technique introduced by \cite{Brunton16} --- was extended by \cite{Kang25} to discover symbolic SGS closure models that recover physically meaningful forms. Sparse SGS closures have also been developed for geophysical turbulence. Using relevance vector machines, \cite{Zanna20} developed closures for ocean mesoscale models and \cite{Jakhar24} for natural convection. To address \textit{a posteriori} stability challenges observed in these works, \cite{Ross23} introduced manual intervention into a hybrid learning framework that combined genetic programming with linear regression, benchmarking the resulting models in idealized ocean simulations. An alternative hybrid strategy has been explored for the integration of neural networks with symbolic regression: \cite{Li24} introduced gene expression networks (GENets) for efficient and robust convergence to explicit turbulence closures models.

The majority of the existing sparsity- and symbolic regression-based RANS and LES closure modeling approaches posed in the literature are constructed with the intention of satisfying the fundamental invariance properties, providing a principled pathway toward model generalization \citep{Ling16}. Nonetheless, evidence of reliable \textit{a posteriori} performance in flow regimes distinct from those used for training is often limited or absent. In some studies, validation is restricted to \textit{a priori} settings \citep{Chakrabarty21, Chung22, Kang25}, where strong performance is neither indicative of numerical stability nor necessarily correlated with \textit{a posteriori} accuracy \citep{Hammond22}. Among those that do undergo \textit{a posteriori} evaluation, generalization is commonly demonstrated only for flows closely aligned with the training data \citep{Reissmann21, Li21, Li24}, often leaving robustness to changes in physics, geometry, or grids untested or unreported. In several cases, these data-driven sparse algebraic closures exhibit numerical instabilities in \textit{a posteriori} \citep{Zanna20, Jakhar24}. Beyond these predictive challenges, many current frameworks require computationally expensive training procedures --- particularly those that use iterative loops involving a CFD solver \citep{Zhao20, Reissmann21,  Lav21, Waschkowski22, BenHassanSaidi22, Xie22, Liao25}, employ non-deterministic training methodologies that require repeated regression passes during model selection \citep{Weatheritt16, Weatheritt17, Chakrabarty21, Li21, Kang25}, or adopt hybrid or deep learning-based model classes \citep{Cherroud22, Tang23a, Li24, Alhafiz25, Wu23, Wu25}. These burdens not only diminish the practical efficiency gains that such closure frameworks are meant to offer, but may also be unnecessary given that sparse models should be learnable from limited, sparse, or noisy data \citep{Brunton24}.

These limitations highlight the need for turbulence closures that retain a structurally transparent form amenable to systematic evaluation of individual term contributions, remain stable and predictive in deployment across flow regimes distinct from those used for training, and maintain efficiency in both the training and inference phases. To address these challenges in the LES context, we introduce a sparsity-promoting SGS closure modeling framework that leverages sparse regression to identify explicit algebraic model forms while retaining only the most physically relevant terms. The framework not only enforces fundamental invariance and symmetry properties, but also embeds arbitrary filter anisotropy and penalizes SGS dissipation. We embed symmetry, Galilean invariance, unit invariance, reflectional invariance, and rotational invariance by training over a minimal tensor basis that is scaled by a truncated infinite polynomial expansion of invariant scalars. This is executed in a manner that carefully regulates the highest order of the candidate model terms. Arbitrary filter anisotropy is embedded in the closure model form by utilizing a mapping between the physical space --- with an anisotropic filter kernel --- and the parent space --- with an isotropic filter kernel --- as introduced in \cite{Prakash24}. This is an important feature that existing data-driven SGS closure formulations often inadequately address or overlook; filter anisotropy affects the representation of turbulent structures at different scales and can influence the accuracy of SGS models in capturing flow dynamics. As such, explicitly accounting for filter anisotropy enables models to represent turbulent structures that span different scales and grid resolutions in different directions, improving robustness and enabling more reliable \textit{a posteriori} predictions across diverse flow configurations \cite{Prakash24}. Another critical aspect commonly overlooked by data-driven SGS closures is the accurate representation of SGS energy dissipation. Modeling approaches that seek to reconstruct the components of the SGS tensor often underpredict the SGS energy dissipation \citep{Vreman97}, which can compromise numerical stability. To address this, we implement a tailored objective function that penalizes dissipation errors during training. These considerations of sparsity, invariance properties, filter anisotropy, and dissipation correction collectively reduce reliance on memory- and compute-intensive training procedures, enabling robust and generalizable SGS model discovery from small training datasets. 

The remainder of this paper is structured as follows. In Section~\ref{sec:LES}, we introduce the filtered Navier-Stokes Equations that govern LES and discuss the associated closure problem. In Section~\ref{sec:develop}, we detail the development of the SGS modeling framework, including the choice of model dependencies (\ref{sec:inputs}), the generalization to arbitrary grid anisotropy (\ref{sec:aniso}), the embedding of invariance properties (\ref{sec:invariance}), the sparse regression algorithm (\ref{sec:SR}), and the modulation of the model dissipation error (\ref{sec:model dissip}). In Section~\ref{sec:results}, we describe the model training procedure (\ref{sec:training}) before evaluating the predictive and computational performance of the modeling framework. Predictive performance is assessed by examining how parameter choices influence learned sparse regression closures under both \textit{a priori} (\ref{sec:model order}) and \textit{a posteriori} (\ref{sec:penalty parameters}) conditions across diverse flow regimes, including forced homogeneous isotropic turbulence (HIT), Taylor–Green vortex flow, and a turbulent channel flow. The periodic hill benchmark (\ref{sec:periodic hill}) is then used to assess the generalizability of the framework to separated flows. This is followed by a computational efficiency benchmark (\ref{sec:efficiency}) to quantify the training and inference costs. To conclude, in Section~\ref{sec:conclusions}, we summarize the modeling framework's features, emphasize key results and takeaways, and suggest potential avenues for future research.

%% ---------------------------
%% Large Eddy Simulation
%% ---------------------------
\section{The Large Eddy Simulation Closure Problem} \label{sec:LES}

While the Navier-Stokes equations can, in principle, resolve the full spectrum of turbulence scales via direct numerical simulation (DNS), the associated computational expense is prohibitive for most practical flows. Instead, LES maintains tractability for a broader range of applications by solving the filtered Navier-Stokes equations. Assuming a homogeneous filter, the filtered incompressible Navier-Stokes equations take the form:

\begin{equation}
    \frac{\partial\bar{u}_{i}}{\partial x_{i}} = 0,
    \label{eqn:filtered continuity}
\end{equation}

\begin{equation}
    % time derivative
    \frac{\partial \bar{u}_{i}}{\partial t} + 
        % advection
        \frac{\partial}{\partial x_{j}} \left( \bar{u}_{i}\bar{u}_{j} \right) = 
        % pressure gradient
        -\frac{1}{\rho}\frac{\partial \bar{p}}{\partial x_{i}} 
        % viscous forces
        + \frac{\partial}{\partial x_{j}} ( 2 \nu \bar{S}_{ij}) 
        % SGS stress
        - \frac{\partial \tau_{ij}}{\partial x_{j}} +
        % body forces
        \bar{f}_{i},
    \label{eqn:filtered NSE}
\end{equation}

\noindent where $\bar{u}_{i}$ is the $i^{th}$ component of the filtered velocity, $\rho$ is the fluid density, $\bar{p}$ is the filtered pressure, $\nu$ is the kinematic viscosity of the fluid, $\bar{S}_{ij} = \frac{\mathrm{1}}{\mathrm{2}} (\partial \bar{u}_i/\partial x_j + \partial \bar{u}_j/\partial x_i)$ is the $ij^{th}$ component of the filtered strain rate tensor, $\tau_{ij} = \overline{u_iu_j} - \bar{u}_i\bar{u}_j$ is the $ij^{th}$ component of the SGS stress tensor, and $\bar{f}_{i}$ is the $i^{th}$ component of the body force. 

The filtering technique applied to the Navier-Stokes equations aims to reduce the complexity of the numerical solution by retaining only the physics of the large scales, resulting in the effective loss of information regarding the small scales. As a consequence, the variable that is constructed from subgrid-scale-dependent terms --- $\tau_{ij}$ --- cannot be calculated directly. This leaves the filtered Navier-Stokes equations unclosed, as there are more unknown variables than there are equations. This issue is known as the closure problem in LES. 

In a turbulent flow, turbulent kinetic energy is transferred predominantly from large to small scales. However, on local scales, the transfer of energy is bidirectional \citep{Kraichnan75, Pouquet19}; that is, energy is exchanged between scales via both forwardscatter (corresponding to the classical energy cascade that represents the energy transport from large to small scales) and backscatter (corresponding to the inverse energy cascade that represents the energy transport from small to large scales). These complex energy transfer dynamics can be distorted if $\tau_{ij}$ is not modeled accurately. Furthermore, it is essential for an SGS closure model to reflect the unresolved interactions between the large and small scales to an appropriate degree of accuracy, as the reliability of any LES is contingent upon its model's fidelity \citep{Sagaut06}. 

There are two predominant approaches to modeling the SGS stress tensor: functional and structural. The functional approach aims to forecast the effect of the subgrid terms on the resolved variables by predicting the SGS energy dissipation. This inherently dissipative formulation enforces a forward energy cascade, resulting in purely diffusive models that promote numerical stability \citep{Sagaut06} but suppress backscatter and exhibit poor correlation with the exact SGS tensor \citep{Borue98}. The structural approach, on the other hand, aims to reconstruct the SGS tensor components from resolved quantities, invoking an assumption of scale similarity between the resolved and unresolved structures. Although structural models often yield strong \textit{a priori} correlations with the exact SGS tensor \citep{Borue98}, they tend to underpredict the SGS energy dissipation, commonly compromising \textit{a posteriori} performance \citep{Vreman97} and numerical stability. For a more thorough review of functional and structural SGS stress modeling approaches and preexisting models, the reader is referred to \cite{Sagaut06}.

%% ---------------------------
%% Development of the Modeling Framework
%% ---------------------------
\section{Development of the Modeling Framework}\label{sec:develop}

Our systematic process of developing a data-driven SGS closure modeling framework begins with the preselection of an appropriate model form and structure. The framework developed herein is designed to yield SGS closure forms that preserve the fundamental symmetry and physical invariance properties of the SGS stress tensor, ensuring generalizability and robustness, while incorporating filter anisotropy to broaden applicability to flow configurations with anisotropic grids. The SGS closure structure adopted here is that of explicit algebraic nonlinear eddy viscosity (NLEV) models. In this class of models, the deviatoric component of the SGS stress tensor, $\tau_{ij}^d = \tau_{ij} - (1/3)\tau_{kk} \delta_{ij}$, is expressed as a linear combination of products between scaling coefficients $g^{(n)}$ and basis tensors $T_{ij}^{(n)}$:

\begin{equation}
    \tau_{ij}^d = \sum_{n=0}^{N} g^{(n)} T_{ij}^{(n)} .
    \label{eqn:eddy viscosity model}
\end{equation}

\noindent Some formulations instead model the full SGS stress tensor. In the case of LEV models, $N=\mathrm{0}$, $g^{(\mathrm{0})}=-\mathrm{2}\nu_t$, and $T_{ij}^{(\mathrm{0})}=\bar{S}_{ij}$, giving $\tau_{ij}^d=-\mathrm{2}\nu_{t}\bar{S}_{ij}$. In contrast, NLEV models retain the general form posed in Equation~\eqref{eqn:eddy viscosity model} and include nonlinear basis tensors. In our framework, the scaling coefficients are determined via sparse regression, yielding parsimonious algebraic closures. We aim to improve the functional performance of the yielded structural models by utilizing a tailored objective function that penalizes errors in SGS dissipation during the training phase.

In this section, we provide a detailed outline of our approach to developing the modeling framework, enabling replication and further extension of our work. The selection criteria for our model dependencies are discussed first in Section~\ref{sec:inputs}. This is followed by a description of how we generalize to arbitrary filter anisotropy in Section~\ref{sec:aniso}. Next, we explain how to embed physical invariance properties directly into the model form in Section~\ref{sec:invariance}. We then demonstrate how the sparse regression algorithm is implemented to enable user optimization and customization in Section~\ref{sec:SR}. Lastly, in Section~\ref{sec:model dissip}, we present our functionally informed objective function aimed at reducing the SGS energy dissipation error.

%% ---------------------------
%% Development: Model Dependencies
%% ---------------------------
\subsection{Choice of Model Dependencies} \label{sec:inputs}

A key component in the development of a data-driven closure modeling framework is the selection of resolved flow field variables on which the SGS tensor is assumed to depend. To do this, we follow the same approach set forth by \cite{Prakash22, Prakash24} and outline a series of constraints on the inputs of the closure modeling framework. Accordingly, the inputs must:

\begin{enumerate}

  \item be local in space and time to yield computationally efficient models;
  
  \item preserve Galilean invariance in the yielded models by having dependence on quantities that are themselves Galilean invariant; 

  \item distinguish different local flow states to yield SGS models that correlate to different values of the local derivatives of the filtered velocity field;

  \item recognize different filter widths and aspect ratios to yield SGS models that correlate to different filter length scales and anisotropies;

  \item account for near-wall behavior by having dependence on viscosity to properly turn off the yielded models in over-resolved wall boundary regions of the flow.
  
\end{enumerate}

\noindent The dimensional model form that results from the fulfillment of these requirements by a minimal set of inputs is expressed as follows:

\begin{equation}
    \tau_{ij}=\tau_{ij}^\mathrm{model}(\bar{\boldsymbol{G}}, \boldsymbol{A}, \varDelta, \nu)
    \label{eqn:physical space dimensional model form}
\end{equation}

\noindent where $\bar{\boldsymbol{G}} = \bar{G}_{ij} = \partial \bar{u}_i / \partial x_j$ is the velocity gradient tensor, $\boldsymbol{A}$ is the so-called anisotropy tensor (to be defined thoroughly in the following section) that characterizes the shape of the filter kernel, and $\varDelta$ characterizes the size of the filter kernel.

%% ---------------------------
%% Development: Anisotropy
%% ---------------------------
\subsection{Generalizing to Anisotropic Grids}\label{sec:aniso}

To reduce the computational overhead in wall-bounded turbulent flows, anisotropic grid resolutions are commonly employed to align grid spacing with boundary layer structures. In such settings, arbitrary filter anisotropy should be properly accounted for. Deconvolution-based SGS models inherently incorporate filter anisotropy via exact \citep{Germano86a, Germano86b, Bull16} or approximate \citep{Stolz99, Stolz01a, Stolz01b} invertible filtering operations. Eddy viscosity closure formulations do not; therefore, anisotropic grid effects are often inadequately represented or entirely unaccounted for \citep{Haering19, Schumann20, Moser21, Prakash23}. 

In the data-driven turbulence modeling community, anisotropic resolutions are frequently treated heuristically by introducing scalar characteristic filter widths (e.g., the geometric mean, maximum, or root mean square of the filter width components) as model inputs \citep{Zhou19, Xie20, Reissmann21, Prakash22, Lee23, Wu24}. Such scalar measures, however, cannot uniquely characterize an arbitrary anisotropic grid \citep{Haering19, Schumann20, Prakash24, Prakash23}. An alternative approach is to train closure models on anisotropically filtered DNS (fDNS) data \citep{Prakash24, Meng23}. This approach is employed by \cite{Meng23} to realize SGS stress and heat flux models for a compressible turbulent channel flow with the use of fully connected neural networks. The SGS models were assessed in both \textit{a priori} and \textit{a posteriori} tests, albeit only for a single flow configuration; it is reasonable to expect that these models would not extend well to flow configurations that the models were not trained on given that no physical invariance properties are embedded in the model form. Furthermore, \textit{a priori} tests revealed that the models set forth by \cite{Meng23} suffer from reduced predictive capabilities when tested on grid anisotropies outside of the training dataset, suggesting a lack of generalizability to arbitrary filter anisotropy. 

\begin{figure}[b]
  \centerline{\includegraphics[width=\textwidth]{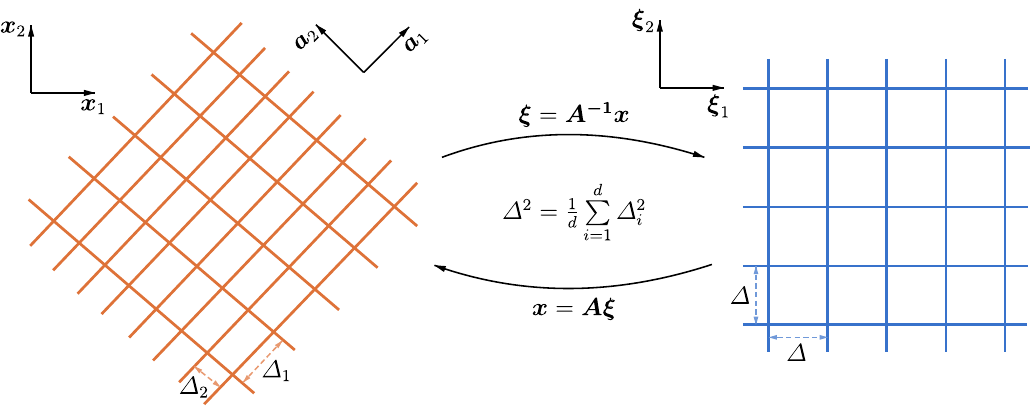}}% Images in 100% size
  \caption{Mapping between an anisotropic grid in the physical filtering space and an isotropic grid in the parent filtering space \citep{Prakash24}.}
    \label{fig:anisotropic grid mapping}
\end{figure}

As demonstrated by \cite{Prakash24}, filter anisotropy can be embedded directly into the closure model formulation through a forward linear mapping $\boldsymbol{\mathit{\xi}}:\mathbb{R}^3 \rightarrow \mathbb{R}^3$ that transforms spatial coordinates between the physical anisotropic filter space --- with coordinates $\boldsymbol{x}$ --- and a parent isotropic filter space:

\begin{equation}
    \boldsymbol{\mathit{\xi}}(\boldsymbol{x}) =
        \boldsymbol{A}^{-1}
        \boldsymbol{x}.
    \label{eqn:forward mapping}
\end{equation}

\noindent Figure~\ref{fig:anisotropic grid mapping} shows a visual representation of this forward mapping and its associated inverse $\boldsymbol{x}(\boldsymbol{\xi}) = \boldsymbol{A \xi}$, where $\boldsymbol{A}$ is the so-called anisotropy tensor in a $d$-dimensional filtering space:

\begin{equation}
    A_{ij} = 
        \sqrt{d}
        \frac{\varDelta_{ij}} {\|\boldsymbol{\varDelta}\|_F}.
    \label{eqn:A}
\end{equation}

\noindent In the above, $||\boldsymbol{\cdot}||_F$ denotes the Frobenius norm and $\boldsymbol{\varDelta} = \sum_{i=1}^{d} \varDelta_i\, \boldsymbol{a}_i \otimes \boldsymbol{a}_i$ is the filter width tensor, which parameterizes the filtering domain. The filter width vector $\varDelta_i$ designates the size of the filter width in each of the principal filtering directions $\boldsymbol{a}_i$. For isotropic filters, $\boldsymbol{A}=\boldsymbol{I}$. Note that here, isotropy and anisotropy refer specifically to the shape of the filter kernel --- spherical in isotropic filtering domains and ellipsoidal in anisotropic filtering domains --- not to the spatial variation of the filter width. The latter can give rise to commutation errors in the SGS term. For instances, however, where the filter width varies spatially and the anisotropy tensor cannot be prescribed globally, \ref{sec:appendix -- aniso local} provides a local definition of $\boldsymbol{A}$ in which the filtering geometry is defined according to the local geometry of the underlying computational grid.

We utilize this same approach to embed filter anisotropy in the modeling framework at hand. A concise overview of the underlying mathematics is provided here, however the reader is referred to \cite{Prakash24} for its rigorous theoretical foundation. The adoption of this mapping enables us to represent our filtered flow variables in the physical space --- denoted by $\left(\,\mathcal{\bar{\cdot}}\,\right)$ --- as filtered variables in the parent space --- denoted by $\left(\,\mathcal{\tilde{\cdot}}\,\right)$. Accordingly, the SGS tensor in the physical space can be related to an analogous tensor in the parent space, yielding the SGS tensor anisotropy identity:

\begin{equation}
    \overline{u_i u_j}(\boldsymbol{x}) - \bar{u}_i(\boldsymbol{x}) \bar{u}_j(\boldsymbol{x}) =
        \widetilde{(u_i \circ \boldsymbol{x}) (u_j \circ \boldsymbol{x})}
            (\boldsymbol{\mathit{\xi}}(\boldsymbol{x})) -
        (\widetilde{u_i \circ \boldsymbol{x}})(\boldsymbol{\mathit{\xi}}(\boldsymbol{x})) \;
        (\widetilde{u_j \circ \boldsymbol{x}})(\boldsymbol{\mathit{\xi}}(\boldsymbol{x})).
    \label{eqn:SGS mapping}
\end{equation}

\noindent With the use of this identity, the model form of the SGS tensor written in Equation~\eqref{eqn:physical space dimensional model form} can be expressed in terms of the flow variables in the mapped isotropic space, giving:

\begin{equation}
    \tau_{ij}
        = \tau_{ij}^\mathrm{model}(\tilde{\boldsymbol{G}}, \boldsymbol{I}, \varDelta, \nu)
        = \tau_{ij}^\mathrm{model}(\tilde{\boldsymbol{G}}, \varDelta, \nu),
    \label{eqn:parent space dimensional model form}
\end{equation}

\noindent where $\tilde{\boldsymbol{G}}=\tilde{G}_{ij}=\partial(\widetilde{u_i \circ x})/\partial \xi_j$ is the gradient of the filtered velocity in the parent isotropic filter space. $\tilde{G}_{ij}$ is related to $\bar{G}_{ij}$ via the following relation:

\begin{equation}
    \tilde{G}_{ij}
        =\frac{\partial(\widetilde{u_i \circ x})}{\partial \xi_j} 
        =\frac{\partial\left(\bar{u}_i \circ x\right)}{\partial x_k} 
            \frac{x_k}{\partial \xi_j}
        =\frac{\partial \bar{u}_i}{\partial x_k} 
            \frac{x_k}{\partial \xi_j}
        =\bar{G}_{ik}  A_{kj} .
    \label{eqn:G mapping}
\end{equation}

\noindent This definition enables the representation of the symmetric $\tilde{S}_{ij}$ and anti-symmetric $\tilde{\varOmega}_{ij}$ parts of the gradient of the filtered velocity in the parent filter space as

\begin{equation}
    \tilde{S}_{i j}
        =\tilde{G}_{i j}^{\,\mathrm{sym}}
        =\frac{\tilde{G}_{i j}+\tilde{G}_{j i}}{2}
    \label{eqn:S mapping}
\end{equation}

\noindent and

\begin{equation}
    \tilde{\varOmega}_{i j}
        =\tilde{G}_{i j}^{\,\mathrm{anti}-\mathrm{sym}}
        =\frac{\tilde{G}_{i j}-\tilde{G}_{j i}}{2},
    \label{eqn:Omega mapping}
\end{equation}

\noindent which reduce to the standard filtered strain-rate $\bar{S}_{ij}= \tilde{\boldsymbol{S}}$ and rotation-rate $\bar{\varOmega}_{ij}= \tilde{\boldsymbol{\varOmega}} =\frac{\mathrm{1}}{\mathrm{2}} (\partial \bar{u}_i/\partial x_j -$ $\partial \bar{u}_j/\partial x_i)$ on an isotropic grid. We can also attain the magnitude of the gradient of the filtered velocity in the mapped space $\tilde{G}$:

\begin{equation}
    \tilde{G}
        =\left( \tilde{S}_{ij} \tilde{S}_{ij} + \tilde{\varOmega}_{ij} \tilde{\varOmega}_{ij} \right)^{1/2}.
    \label{eqn:G magnitude}
\end{equation}

\noindent The final dimensional model form of the SGS tensor resulting from these definitions is

\begin{equation}
    \tau_{ij}=\tau_{ij}^\mathrm{model}(\tilde{\boldsymbol{S}}, \tilde{\boldsymbol{\varOmega}}, \tilde{G}, \varDelta, \nu).
    \label{eqn:final dimensional model form}
\end{equation}

The approach outlined above to embed arbitrary filter anisotropy directly into the model form of SGS closures has already been implemented for a data-driven SGS model in the filtered strain-rate eigenframe ($\tilde{S}$-frame) \citep{Prakash24} and for the commonly used Smagorinsky model \citep{Prakash23}. \cite{Prakash24} demonstrated the ability of the anisotropic $\tilde{S}$-frame model to generalize to several flow configurations and grid resolutions outside of the dataset used to train the model. The authors believe that this approach to embedding filter anisotropy largely contributed to the model's predictive capabilities, as it consistently outperformed other state-of-the-art SGS models in both \textit{a priori} and \textit{a posteriori} tests. The static Smagorinsky model with embedded anisotropy \citep{Prakash23} attained similar levels of success in both \textit{a priori} and \textit{a posteriori} tests for a variety of grid resolutions, types, and anisotropies; the anisotropic form routinely performed better than the classical Smagorinsky model with commonly used scalar anisotropy-characterizing length scale definitions. We expect that the extension of this approach to our data-driven NLEV closure modeling framework will yield SGS models that exhibit features and functionalities similar to those in \cite{Prakash24, Prakash23} but with the added benefit of an explicit algebraic form. To the best of our knowledge, no existing data-driven sparsity-inducing turbulence closure modeling methodologies explicitly incorporate grid anisotropy into their model formulation.

% \newpage

%% ---------------------------
%% Development: Invariance
%% ---------------------------
\subsection{Embedding Invariance Properties} \label{sec:invariance}

A predominant objective of the modeling framework is to formulate closure models that are generalizable. To accomplish this, it is crucial that the modeled tensors possess the same mathematical properties inherent to true SGS stress tensors --- symmetry, Galilean invariance, unit invariance, reflectional invariance, and rotational invariance. The improved functionality of data-driven models that preserve these properties was first demonstrated for closure of the RANS equations by \cite{Ling16}. In this work, a Reynolds stress closure model with a novel invariance-preserving learning architecture outperformed existing data-driven and classical turbulence models. Most profoundly, the model proposed in \cite{Ling16} yielded results with superior predictive accuracy when tested in both the \textit{a priori} and \textit{a posteriori} on flow physics and configurations outside the training cases. The model's ability to extrapolate and generalize to previously unseen flow cases is attributed to its embedded invariance properties. The introduction of this finding to the data-driven turbulence modeling community has been transformative, as the efficacy of symmetry and invariance-preserving learning methods are now widely recognized \citep{Duraisamy19, Brunton20, Beck21, Vinuesa22, Girimaji24, Sanderse25}. In this section, we outline how Galilean, unit, reflectional, and rotational invariance properties are embedded within the form of our modeling framework.

We begin with Galilean invariance. In Section~\ref{sec:inputs}, the second constraint we placed on our model inputs was that they must be Galilean invariant themselves in order to yield Galilean invariant models. This indicates that SGS models cannot depend on the local filtered velocity field itself, but instead on the spatial derivatives of the filtered velocity field, as $\tilde{S}_{ij}$, $\tilde{\varOmega}_{ij}$, and $\tilde{G}$ do. Furthermore, $\varDelta$ --- which is predetermined by the grid --- and $\nu$ --- which is independent of motion for an incompressible fluid --- are, by definition, Galilean invariant. Therefore, the model form expressed in Equation~\eqref{eqn:final dimensional model form} already represents a Galilean invariant SGS stress model. 

Next, to attain a unit invariant model form, we nondimensionalize our model inputs and outputs by invoking the Buckingham Pi theorem. The model form in Equation~\eqref{eqn:final dimensional model form} has six variables and two independent dimensions, indicating that there are four independent dimensionless parameters --- denoted using $\left(\mathcal{\,\hat{\cdot}}\,\right)$ --- selected as follows:

\noindent
\begin{minipage}{0.24\textwidth}
    \begin{align}
        &\hat{\tilde{S}}_{ij} = \frac{\tilde{S}_{ij}}{\tilde{G}}, \label{eqn:nondim S}
    \end{align}
\end{minipage}
\hfill
\begin{minipage}{0.24\textwidth}
    \begin{align}
        &\hat{\tilde{\varOmega}}_{ij} = \frac{\tilde{\varOmega}_{ij}}{\tilde{G}}, \label{eqn:nondim omega}
    \end{align}
\end{minipage}
\hfill
\begin{minipage}{0.24\textwidth}
    \begin{align}
        &\hat{\nu} = \frac{\nu}{\varDelta^2 \tilde{G}}, \label{eqn:nondim nu}
    \end{align}
\end{minipage}
\hfill
\begin{minipage}{0.24\textwidth}
    \begin{align}
        &\hat{\tau}_{ij} = \frac{\tau_{ij}}{\varDelta^2 \tilde{G}^2}, \label{eqn:nondim tau}
    \end{align}
\end{minipage}\bigskip
\label{eqn:nondim}

\noindent Taken together, these parameters yield the nondimensional form of the modeled SGS stress tensor:

\begin{equation}
    \tau_{ij} = 
        \varDelta^2 \tilde{G}^2 \hat{\tau}_{ij}^\mathrm{model}(\hat{\tilde{\boldsymbol{S}}}, \hat{\tilde{\boldsymbol{\varOmega}}}, \hat{\nu}).
    \label{eqn:nondim model form}
\end{equation}

Finally, to embed rotational and reflectional invariance into our model form, we assume that the SGS stresses lie on an integrity basis of invariant scalars and tensors (as was done in \cite{Ling16}). Under this assumption, the modeled SGS stress tensor mimics the structure of an NLEV:

\begin{equation}
    \hat{\tau}_{ij}^\mathrm{model} = 
    \sum_{n=0}^{N} 
    g^{(n)}(\lambda_0, ..., \lambda_m)
    T_{ij}^{(n)},
    \label{eqn:tau model NLEV}
\end{equation}

\noindent where the scaling coefficients in this case are functions of scalar invariants $\lambda_0, ..., \lambda_m$. For a prescribed set of input tensors, its integrity basis is discovered by invoking the Cayley-Hamilton theorem to reduce an infinite polynomial expansion of the input tensors to a finite linear combination of basis tensors. The generalized tensor representation theory derived by G.F. Smith in 1971 \citep{Smith71} can be used to obtain an integrity basis to characterize $\hat{\tau}_{ij}$ in terms of the $\hat{\tilde{S}}_{ij}$ and $\hat{\tilde{\varOmega}}_{ij}$ tensors. Using Smith's tensor representation for Equation~\eqref{eqn:tau model NLEV}, $N=\mathrm{7}$, giving eight basis tensors when modeling the full SGS tensor:

\begin{subequations}
    \noindent
    \begin{minipage}{0.48\textwidth}
        \begin{align}
            &\boldsymbol{T}^{(0)} = \boldsymbol{I} \label{eqn:T0} \\
            &\boldsymbol{T}^{(1)} = \hat{\tilde{\boldsymbol{S}}} \label{eqn:T1} \\
            &\boldsymbol{T}^{(2)} = \hat{\tilde{\boldsymbol{S}}}^2 \label{eqn:T2} \\
            &\boldsymbol{T}^{(3)} = \hat{\tilde{\boldsymbol{\varOmega}}}^2 \label{eqn:T3}
        \end{align}
    \end{minipage}
    \hfill
    \begin{minipage}{0.48\textwidth}
        \begin{align}
            &\boldsymbol{T}^{(4)} = \hat{\tilde{\boldsymbol{S}}}\hat{\tilde{\boldsymbol{\varOmega}}} -   
                \hat{\tilde{\boldsymbol{\varOmega}}}\hat{\tilde{\boldsymbol{S}}} \label{eqn:T4} \\
            &\boldsymbol{T}^{(5)} = \hat{\tilde{\boldsymbol{\varOmega}}} \hat{\tilde{\boldsymbol{S}}}        
                \hat{\tilde{\boldsymbol{\varOmega}}} \label{eqn:T5} \\
            &\boldsymbol{T}^{(6)} = \hat{\tilde{\boldsymbol{S}}}^2 \hat{\tilde{\boldsymbol{\varOmega}}} -   
                \hat{\tilde{\boldsymbol{\varOmega}}} \hat{\tilde{\boldsymbol{S}}}^2 \label{eqn:T6} \\
            &\boldsymbol{T}^{(7)} = \hat{\tilde{\boldsymbol{\varOmega}}} \hat{\tilde{\boldsymbol{S}}} 
                \hat{\tilde{\boldsymbol{\varOmega}}}^2 - \hat{\tilde{\boldsymbol{\varOmega}}}^2 
                \hat{\tilde{\boldsymbol{S}}} \hat{\tilde{\boldsymbol{\varOmega}}} \label{eqn:T7}
        \end{align}
    \end{minipage}\bigskip
    \label{eqn:Ts}
\end{subequations}

\noindent and $m=\mathrm{5}$, giving six scalar invariants when viscosity is included as a model input:

\begin{subequations}
    \noindent
    \begin{minipage}{0.48\textwidth}
        \begin{align}
            &\lambda_0 = \mathrm{\mathit{Tr}}(\hat{\tilde{\boldsymbol{S}}}^2) \label{eqn:lambda0} \\
            &\lambda_1 = \mathrm{\mathit{Tr}}(\hat{\tilde{\boldsymbol{\varOmega}}}^2) \label{eqn:lambda1} \\
            &\lambda_2 = \mathrm{\mathit{Tr}}(\hat{\tilde{\boldsymbol{S}}}^3) \label{eqn:lambda2}
        \end{align}
    \end{minipage}
    \hfill
    \begin{minipage}{0.48\textwidth}
        \begin{align}
            &\lambda_3 = \mathrm{\mathit{Tr}}(\hat{\tilde{\boldsymbol{\varOmega}}}^2 \hat{\tilde{\boldsymbol{S}}}) \label{eqn:lambda3} \\
            &\lambda_4 = \mathrm{\mathit{Tr}}(\hat{\tilde{\boldsymbol{\varOmega}}}^2 \hat{\tilde{\boldsymbol{S}}}^2) \\
            &\lambda_5 = \hat{\nu}, \label{eqn:lambda5} 
        \end{align}
    \end{minipage}\bigskip
    \label{eqn:lambdas}
\end{subequations}

\noindent where $\mathrm{\mathit{Tr}}$ denotes the trace. Smith's representation theory is known to be complete and minimal \citep{Pennisi87}, meaning that the constructed tensor polynomial is expressed by the smallest possible number of terms without loss of generality \citep{Stallcup22}. 

An alternate representation of the SGS stress tensor in terms of the strain and rotation rate tensors was proposed by S.B Pope in 1975 \citep{Pope75}. Numerous data-driven turbulence models use Pope's integrity basis (and occasionally simplifications of this basis by \cite{Lund93}) for RANS \citep{Ling16, Weatheritt17, Schmelzer20, Waschkowski22, Tang23b} and LES \citep{Xie20, Reissmann21, Li21, Li23, Bose24} closure. Regrettably, Pope's basis is neither minimal nor complete \citep{Stallcup22}, but is still used more widely for data-driven closure modeling than Smith's basis. 

The mathematical foundations used to derive these representation theories --- Hilbert's Basis Theorem, the Cayley-Hamilton Theorem, and procedures documented by Rivlin \citep{Spencer58} --- can be used to extend an integrity basis's dependence on additional tensor quantities. Extended bases have been developed to account for additional variables in data-driven turbulence closure models such as the gradients of pressure \citep{Parmar20} and turbulent kinetic energy \citep{Wu18, Steiner22}. A primary drawback of this extension, however, is that the size of the integrity basis (and its associated computational expense) grows exponentially as the number of tensor inputs increases. For this reason, data-driven NLEV closure models of the form presented here become intractable as additional tensor quantities are considered. An alternative strategy that involves modeling the Reynolds stress \cite{Peters22} or SGS stress \cite{Prakash22} eigenstructure eliminates this impediment entirely. Nevertheless, the approach outlined above is sufficient for the SGS modeling framework developed herein, as we only need two tensor quantities to satisfy the series of constraints placed on our model inputs in Section~\ref{sec:inputs}.

% \newpage

%% ---------------------------
%% Development: Sparse Regression
%% ---------------------------
\subsection{Machine Learning Algorithm} \label{sec:SR}

The goal of the machine learning algorithm is to determine the scaling coefficients, as the tensor basis (Equation~\eqref{eqn:Ts}) and the scalar invariants (Equation~\eqref{eqn:lambdas}) are known \textit{a priori}. Data-driven approaches to finding these scaling coefficients often involve using neural networks \citep{Ling16, Parmar20, Xie20, Man23, Tang23b, Wu24, Bose24}. The most common approach to ensuring that invariance properties are preserved when using neural networks to learn SGS stress models is to use an invariance-preserving architecture known as a tensor basis neural network (TBNN), which was first introduced by \cite{Ling16}. TBNNs differ from standard feed-forward multilayer perceptron (MLP) neural networks by having two input layers: an invariant input layer for the scalar invariants and a tensor input layer for the tensor basis. Once the scalar invariants are fed into the invariant input layer, they are passed through a series of hidden layers, where the final hidden layer has one node per tensor within the tensor basis. Each node within this final hidden layer represents a scaling coefficient. As such, in a merge layer, each of the tensors passed through their designated node in the tensor input layer are scaled by their corresponding node in the final hidden layer, then the resulting scaled tensors are summed component-wise to yield the modeled SGS or Reynolds stress tensor. Turbulence modeling methods that utilize TBNNs have demonstrated their ability to yield accurate predictions \citep{Ling16, Parmar20, Xie20, Man23, Tang23b, Wu24, Bose24}. Neural networks, however, often lack interpretability due to their “black box” nature, which inevitably obscures the underlying physics and routinely increases the computational cost over that of classical models. Moreover, neural network-based turbulence models often need to be trained on extensive libraries of diverse datasets to adequately represent the underlying physics, generalize well, and prevent overfitting. The increasing accessibility of data enables such approaches. Nevertheless, the costs associated with acquiring, processing, and storing these comprehensive high-fidelity libraries can be prohibitive.

To address the issues that plague neural networks, our data-driven modeling framework uses sparse regression to discover explicit algebraic NLEV models of the SGS stress tensor by penalizing large models so that only the dominant variables that most represent the flow physics are selected. Regression algorithms in machine learning aim to characterize the general relationship between a response variable and its assumed predictors. In our case, the response variable is $\hat{\tau}_{ij}$ and the assumed predictors are functions of $\lambda_\mathrm{0}, ..., \lambda_m$ and $T_{ij}^{(n)}$. 

We formulate the predictors by scaling Smith's complete and minimal tensor basis by a truncated infinite polynomial expansion of the invariant scalars. This is accomplished by separately grouping the invariant scalars and basis tensors into sets based on their degree $k$, denoted by $\Lambda^{\{k\}}$ and $\Tau^{\{k\}}$, respectively. Predictor sets of a given degree, denoted $P^{\{k\}}$, are then constructed by appropriately scaling basis tensor sets with invariant scalar set products of combined degree $k$. The predictor set for degree $0$, for example, consists of degree-zero tensors scaled by degree-zero scalars. Predictor sets of higher degree are formed as unions of all scalar and tensor set products whose individual degrees sum to the desired total degree:

\begin{equation}
    P^{\{k\}} = \bigcup_{i+j=k} \Lambda^{\{i\}} \, \Tau^{\{j\}}.
    \label{eqn:P union}
\end{equation}

\noindent Explicit degree assignments for the invariant scalars and basis tensors, as well as the resulting predictor set enumerations, are provided in \ref{sec:appendix -- predictor} for reference and reproducibility. 

This degree-based construction enables systematic control over model complexity by truncating the predictor sets at a prescribed maximum degree. The full set of predictors for a model of maximum degree $K$ --- which could, in principle, extend to $\infty$ --- is constructed by taking the union of all single-degree sets (defined in Equation~\eqref{eqn:P union}) up to $K$:

\begin{equation}
    \begin{aligned}
        P^{\{k\leq K\}} (\hat{\tilde{\boldsymbol{S}}}, \hat{\tilde{\boldsymbol{\varOmega}}}, \hat{\nu}) 
            & = P^{\{k=0\}} \cup P^{\{k=1\}} \cup \cdots \cup P^{\{k=K\}}\\
            & = \{ \boldsymbol{P}_n \;|\; n \in N \}
    \end{aligned}
    \label{eqn:PleqK degree set}
\end{equation}

\noindent  where $\boldsymbol{P}_n$ denotes a single predictor in the degree-truncated predictor set $P^{\{k\leq K\}}$ and $N$ is the total number of predictors. For instance, $K=\mathrm{2}$ gives $N=\mathrm{10}$, while $K=\mathrm{3}$ gives $N=\mathrm{23}$ (see \ref{sec:appendix -- predictor}). The modeled SGS stress tensor can be expressed as a linear combination of the predictors:

\begin{equation}
    \hat{\tau}_{ij}^\mathrm{model} (P^{\{k\leq K\}}(\hat{\tilde{\boldsymbol{S}}}, \hat{\tilde{\boldsymbol{\varOmega}}}, \hat{\nu});\boldsymbol{w})
        = \sum_{n \in N } 
            w_n
            \boldsymbol{P}_n.
    \label{eqn:tau model f(P;w)}
\end{equation}

\noindent where $\boldsymbol{w}$ is a vector composed of the coefficients, $w_n$, associated with each predictor $\boldsymbol{P}_n$. 

These coefficients are determined using a regression algorithm that fits a mathematical model to the $\hat{\tau}_{ij}=\sum_{n\in N}w_n\boldsymbol{P}_n$ system using a coordinate descent algorithm such that

\begin{equation}
    \argmin_{\boldsymbol{w}} \, L(\boldsymbol{w}),
    \label{eqn:argmin}
\end{equation}

\noindent where $L(\boldsymbol{w})$ is the objective function. In practice, enforcing $\hat{\tau}_{ij}=\sum_{n\in N}w_n\boldsymbol{P}_n$ over a collection of data samples typically yields an overdetermined system of algebraic equations for $\boldsymbol{w}$, for which an exact solution generally does not exist. As such, the regression problem is formulated to determine $\boldsymbol{w}$ by minimizing the prescribed objective function $L(\boldsymbol{w})$, which provides criteria for selecting an optimal approximation to the overdetermined system. The maximum model order $K$ is an important specification in this regression problem, as increasing $K$ expands the candidate space and raises the cost of identifying an optimal coefficient set. The choice of $K$-value can also affect model robustness upon deployment, as higher-order SGS models are more prone to numerical instabilities in \textit{a posteriori} tests.

For many sparse regression problems, elastic net is the objective function of choice, as it can effectively handle highly correlated predictors without inconsistency or redundancy, prevent both underfitting and overfitting, and perform feature selection. For the system at hand, the elastic net objective function is written as follows:

\begin{equation}
    L(\boldsymbol{w})=
        \underbrace{ \frac{1}{2 M} \sum_{m=1}^M \left[ \hat{\tau}_{ij}^{(m)} - \hat{\tau}_{ij}^{\mathrm{model}(m)} \right]^2} 
            _{\substack{\text{ ordinary least squares } \\ \text{ of SGS stress }}}
        + \underbrace{ \vphantom{\frac{1}{2}} \alpha \eta \sum_{n=1}^N \left| w_n \right|} 
            _{\substack{\text{ Lasso }\vphantom{y} \\ \text{ penalty }}}
        + \underbrace{ \frac{1}{2} \alpha (1-\eta) \sum_{n=1}^N \left( w_n \right)^2} 
            _{\substack{\text{ Ridge } \\ \text{ penalty }}},
    \label{eqn:elastic net}
\end{equation}

\noindent where $M$ is the number of observations, $\alpha$ is the regularization penalty parameter on the model coefficients, and $\mathrm{0}\leq\eta\leq\mathrm{1}$ is the mixing parameter between the Lasso and Ridge penalty terms. The first term --- ordinary least squares of SGS stress --- minimizes the error between the exact ($\hat{\tau}_{ij}$) and predicted ($\hat{\tau}_{ij}^\mathrm{model}$) nondimensional SGS stress tensors, the second term --- Lasso penalty or $L^\mathrm{1}$ norm of the model coefficients --- induces sparsity by preserving only the predictors that are most correlated with $\hat{\tau}_{ij}$, and the third term --- Ridge penalty or $L^\mathrm{2}$ norm of the model coefficients --- promotes regularization by penalizing coefficient magnitudes and mitigating sensitivity to multicollinearity among predictors.

The choice of $\alpha$ in Equation~\eqref{eqn:elastic net} governs the strength of regularization imposed during fitting. Stronger regularization (higher $\alpha$-values) produces sparser models with smaller-magnitude coefficients, which can reduce the computational cost and simplify diagnostic analyses by limiting the number of active terms and isolating their individual contributions. Excessive regularization, however, could lead to underfitting and degraded predictive accuracy. Weaker regularization (lower $\alpha$-values) allows more coefficients to take significant values, which can improve fit to the training data but may also increase model complexity, noise sensitivity, and computational expense. The choice of $\eta$ in Equation~\eqref{eqn:elastic net} sets the ratio of mixing between the Lasso and Ridge penalty terms. If $\eta=\mathrm{0}$, the objective function reduces to Ridge regression, giving a dense $\boldsymbol{w}$. If $\eta=\mathrm{0.5}$, equal weight is given to the Lasso and Ridge penalty terms, yielding  $L^\mathrm{1}$- and $L^\mathrm{2}$-regularized models. If $\eta=\mathrm{1}$, the objective function reduces to Lasso regression, which gives a sparse $\boldsymbol{w}$, but highly correlated predictors may be downselected arbitrarily.

%% ---------------------------
%% Development: Dissipation
%% ---------------------------
\subsection{Modulating Model Dissipation Error} \label{sec:model dissip}

As mentioned in Section~\ref{sec:LES}, structural approaches to modeling the SGS stress tensor are known to suffer from poor functional performance by inaccurately predicting the SGS energy dissipation in \textit{a posteriori} tests, which can lead to instabilities that are commonly addressed using ad-hoc strategies such as clipping \citep{Prakash21}. We take a structural-type approach to modeling the SGS stress tensor, aiming to accurately reconstruct its components using resolved quantities. In an aim to more appropriately balance functional performance and enhance \textit{a posteriori} stability, we modify the elastic net objective function by including an additional term that penalizes the SGS dissipation. Our augmented objective function is expressed as follows:

\begin{multline}
    L(\boldsymbol{w})=
        \frac{1}{2 M} \sum_{m=1}^M 
            \left[ \hat{\tau}_{ij}^{(m)} - \hat{\tau}_{ij}^{\mathrm{model}(m)} \right]^2
        + \underbrace{ 
        \frac{1}{2 M} \beta \sum_{m=1}^M
            \left[ \hat{\tau}_{ij}^{(m)} \hat{\tilde{S}}_{ij}^{(m)} - \hat{\tau}_{ij}^{\mathrm{model}(m)} \hat{\tilde{S}}_{ij}^{(m)} \right]^2 }
            _{\substack{\text{ ordinary least squares } \\ \text{ of SGS dissipation }}} 
    \\ 
         \hphantom{\frac{1}{2}}+ \vphantom{\frac{1}{2}} \alpha \eta \sum_{n=1}^N 
            \left| w_n \right|
        + \frac{1}{2} \alpha (1-\eta) \sum_{n=1}^N 
            \left( w_n \right)^2,
    \label{eqn:custom objective function}
\end{multline} 

\noindent where $\beta$ is the penalty parameter on the ordinary least squares of the SGS dissipation. The choice of $\beta$ weights the relative importance of the SGS dissipation prediction (that is, functional performance) to that of the SGS stress prediction (that is, structural performance). If $\beta=\mathrm{0}$, the elastic net objective function (Equation~\eqref{eqn:elastic net}) is recovered. If $\mathrm{0}<\beta<\mathrm{1}$, the objective places greater emphasis on structural performance. If $\beta=\mathrm{1}$, the functional and structural components are weighted equally. If $\beta>\mathrm{1}$, the balance is shifted more toward functional performance. Greater emphasis on the dissipation constraint can enhance stability and energy balance but may lead to excessive damping that suppresses turbulent fluctuations and distorts resolved-scale dynamics. Conversely, relaxing this constraint may better-preserve resolved-scale fluctuations, but it also raises the risk of under-dissipation, potentially causing energy accumulation at smaller scales that is destabilizing.

Typically, data-driven turbulence models only utilize objective functions that minimize the error between the exact and predicted SGS tensors. To the best of our knowledge, the invariance-preserving convolutional neural networks proposed by \cite{Wu24} are the only other structural-type data-driven SGS stress models that enforce SGS dissipation corrections within the loss function. It was found in \cite{Wu24}, through a series of \textit{a priori} and \textit{a posteriori} tests, that the predictive accuracy of the neural networks is improved when both the SGS stress and SGS dissipation errors are accounted for during training.

% \newpage

%% ---------------------------
%% Results
%% ---------------------------
\section{Numerical Results}\label{sec:results}

Rather than proposing a single, fixed SGS closure, this work focuses on: (i) characterizing the flexibility of the modeling framework itself, (ii) analyzing the behavior of the resulting closures under variations in key parameters, (iii) evaluating their ability to generalize across multiple flow configurations, and (iv) assessing their computational efficiency. To isolate the impact of individual parameter choices, all data-driven SGS models presented in this section follow a common training procedure and are trained on an identical forced HIT DNS dataset, as documented in Section~\ref{sec:training}. Within this controlled setting, we conduct sensitivity analyses using both \textit{a priori} and \textit{a posteriori} evaluation to examine how variations in key modeling and training hyperparameters influence the behavior of the learned closures. Following these studies, we further assess the framework through two benchmark studies: a separated flow configuration to evaluate its generalization capabilities and predictive accuracy, and a computational performance benchmark to quantify training and inference efficiency.

The first sensitivity study is conducted in Section~\ref{sec:model order} using \textit{a priori} analyses to examine the effect of the predictor set maximum order $K$; $K$ governs the number and order of predictor terms considered by the algorithm and, consequently, the potential expressiveness of the resulting closure. This study is performed outside of an LES context to efficiently evaluate the impact of model order, mitigate the stability concerns associated with higher-order models, and determine whether additional complexity yields meaningful performance improvements. In Section~\ref{sec:penalty parameters}, sensitivity studies are performed using \textit{a posteriori} tests to evaluate the influence of two penalty parameters in the training objective function (Equation~\eqref{eqn:custom objective function}): $\alpha$ (Section~\ref{sec:alpha}), which controls the regularization applied to the model coefficients, balancing $L_1$ sparsity promotion and $L_2$ smoothness; and $\beta$ (Section~\ref{sec:beta}), which scales the contribution of the SGS dissipation correction term. For each parameter sweep, we conduct LES simulations of forced HIT, Taylor-Green vortex, and turbulent channel flow. The forced HIT case mirrors the training dataset and provides a controlled baseline to assess how well the closures reproduce the learned turbulence statistics for this idealized, homogeneous, isotropic, statistically stationary flow. The Taylor-Green vortex introduces transitional and unsteady dynamics, providing a setting to evaluate the closures' ability to capture evolving coherent structures and non-equilibrium turbulence. The turbulent channel flow provides a testbed to examine the closures' performance in a wall-bounded environment, using anisotropic grid resolutions to represent the strong shear and near-wall gradients.

In the first benchmark study in Section~\ref{sec:periodic hill}, we employ the periodic hill to assess generalization to a flow characterized by separation. The periodic hill test case provides a robust challenge for turbulence models, as it features complex flow phenomena such as adverse pressure gradients, flow detachment, reattachment, and recirculation zones. Here, we assess the predictive performance of a selected sparse regression closure relative to other commonly used SGS models. In the second benchmark test in Section~\ref{sec:efficiency}, we analyze the computational efficiency of a representative sparse regression closure during both training and inference. This assessment highlights the practical trade-off between model complexity and computational overhead in commonly utilized data-driven closure architectures, where high predictive accuracy often comes at a significant computational cost.

%% ---------------------------
%% Results: Model Training
%% ---------------------------
\subsection{Model Training}\label{sec:training}

% {\parfillskip=0pt
The tailored loss function expressed in Equation~\eqref{eqn:custom objective function} is used to optimize the coefficients of each SGS model generated by the framework. Although hyperparameters vary between models, all are trained on the same dataset, ensuring that differences in performance when conducting sensitivity analyses can be attributed solely to the hyperparameter settings. We model the full SGS stress tensor, so all $\mathrm{6}$ unique components of $\hat{\tau}_{ij}$ are considered. The training dataset is extracted from a single time step of a forced HIT DNS at a Taylor Reynolds number ($Re_{\lambda}$) of 418 \citep{Li08, Perlman07}. All models are trained on the same set of anisotropically filtered, randomly sampled spatial data points. The filter widths span two aspect ratios (AR): one and three. The specific filter widths, along with additional details regarding the training and testing datasets, are provided in Table~\ref{tab:train/test data}. The smallest filter width corresponds to the DNS grid resolution; including this isotropic filter in the training dataset enables the ML algorithm to deactivate the learned closure model when the SGS stress tensor is zero. The remaining filter widths correspond to a book-type filter ($AR = \frac{\Delta_2}{\Delta_1} = \frac{\Delta_3}{\Delta_1} > 1$) with an AR of three, applied at a progressively increasing base filter width ($\Delta_1$). 
% \par}

\begin{wraptable}{r}{4.5cm}
    \centering
    \begin{tabular}{c}
    \hline\hline
        \underline{\textbf{Dataset}}\\ 
            Training/Testing \\ \hline 
        \underline{\textbf{Flow Type}}\\ 
            Forced HIT \\ 
            at $Re_{\lambda}=418$\\ \hline 
        \underline{\textbf{Number of Samples}}\\ 
            209,715/52,429 \\ \hline
        \underline{\textbf{Spatial Locations}}\\ 
            Randomly sampled in \\
            $x_i \in [0.5\pi,\mathrm{1.5}\pi]$ \\ \hline
        \underline{\textbf{Time}}\\ 
            $t=1$ \\ \hline
        \textbf{Filter Width}\\ 
            \underline{$\left( \Delta_1 \times \Delta_2 \times \Delta_3 \right) / \eta$} \\
            $2.2 \! \times \! 2.2 \! \times \! 2.2$ \\
            $2.2 \! \times \! 6.6 \! \times \! 6.6$ \\
            $6.6 \! \times \! 19.8 \! \times \! 19.8$ \\
            $15.4 \! \times \! 46.2 \! \times \! 46.2$ \\
            $28.6 \! \times \! 85.8 \! \times \! 85.8$ \\
            $46.2 \! \times \! 138.6 \! \times \! 138.6$ \\ \hline\hline
    \end{tabular}
    \caption{\centering Training and Testing Data}
    \label{tab:train/test data}
\end{wraptable}

For comparison, results from models generated by the proposed framework are presented alongside those from a TBNN SGS stress model. By definition, TBNNs preserve invariance properties. We embed filter anisotropy in the TBNN using the approach outlined in Section~\ref{sec:aniso} and train it on the same data (listed in Table~\ref{tab:train/test data}). The TBNN has a single hidden layer with 20 neurons, uses a mean squared error loss function to optimize the weights and biases, and employs a leaky ReLU activation function. TBNNs are the most widely used class of invariance-preserving data-driven turbulence models in the literature, often demonstrating improved predictive accuracy relative to classical turbulence models \citep{Ling16, Parmar20, Xie20, Man23, Tang23b, Wu24, Bose24}, albeit at a greater computational expense. As such, TBNNs serve as a natural benchmark for evaluating the performance of the sparse regression models considered here.

%% ---------------------------
%% A Priori Results: Model Order
%% ---------------------------
\subsection{\textit{A Priori} Sensitivity Analysis: Maximum Predictor Set Order ($K$)} \label{sec:model order}

In LES, the mathematical formulation of an SGS model, including its order, shapes not only the representation of the impact of small-scale structures but also the evolution of resolved-scale dynamics. Lower-order SGS models are computationally efficient and simple to implement but may lack accuracy in representing small-scale turbulence characteristics in more complex flow regimes without adequate grid resolutions. Higher-order LES models, on the other hand, often offer improved representations of these complex nonlinear interactions. Still, they come at the expense of increased computational cost and heightened sensitivity to numerical instability. Hence, careful consideration should be given to the implications of an SGS model's order on the trade-off between these properties.

Through $\textit{a priori}$ tests, we examine the impact of varying the maximum degree $K$ of the predictor set $P^{\{k\leq K\}}$ supplied to the sparse regression algorithm. Since increasing $K$ enlarges the candidate tensor library available to the sparse regression, this analysis also assesses the influence of basis size on the resulting SGS model. Predictor set construction is detailed in Section~\ref{sec:SR} and \ref{sec:appendix -- predictor}. \textit{A priori} performance is assessed using two metrics: the correlation coefficient (CC) and the relative error in mean energy flux (REF). The CC, 

\begin{equation}
    \mathrm{CC} = \sum_{i} \sum_{j} 
        \frac{\langle ( \tau_{ij} - \langle \tau_{ij} \rangle ) 
        ( \tau_{ij}^\mathrm{model} - \langle \tau_{ij}^\mathrm{model} \rangle ) \rangle}
        {\langle ( \tau_{ij} - \langle \tau_{ij} \rangle )^2 \rangle^{1/2}
        \langle ( \tau_{ij}^\mathrm{model} - \langle \tau_{ij}^\mathrm{model} \rangle )^2 \rangle^{1/2}},
    \label{eqn:CC}
\end{equation}

\noindent is used as an indicator of the structural accuracy of the modeled SGS stresses, with values closer to one corresponding to a stronger correlation with the exact stresses. The REF,

\begin{equation}
    \mathrm{REF} = \frac{\langle - \tau_{ij}^\mathrm{model} S_{ij} \rangle
        - \langle - \tau_{ij} S_{ij} \rangle}
        {\langle - \tau_{ij} S_{ij} \rangle}.
    \label{eqn:REF}
\end{equation}

\begin{wraptable}{l}{4.5cm}
    \centering
    \begin{tabular}{c}
    \hline\hline
        \underline{\textbf{Dataset}}\\ 
            Validation \\ \hline 
        \underline{\textbf{Flow Type}}\\ 
            Forced HIT \\
            at $Re_{\lambda}=418$\\ \hline 
        \underline{\textbf{Number of Samples}}\\ 
            262,144 \\ \hline
        \underline{\textbf{Spatial Locations}}\\ 
            Uniformly sampled \\
            from $z=\pi$ slice \\ \hline
        \underline{\textbf{Time}}\\ 
            $t=10$ \\ \hline
        \textbf{Filter Width}\\ 
            \underline{$\Delta = \Delta_1 = \Delta_2 = \Delta_3$} \\
            $\Delta \approx 6.6\eta-72.3\eta$ \\
    \hline\hline
    \end{tabular}
    \caption{\centering Validation Data}
    \label{tab:valid data}
\end{wraptable}

\noindent quantifies the dissipative performance of the model, with positive values signifying over-dissipation and negative values suggesting under-dissipation. Together, these metrics provide a preliminary assessment of SGS models. The \textit{a priori} validation dataset (see Table \ref{tab:valid data}) is drawn from the same simulation as the training data \citep{Li08, Perlman07}, but consists of distinct data points and filter widths absent from the training dataset.

\begin{figure}[b]
\centerline{\includegraphics[width=0.99\textwidth]{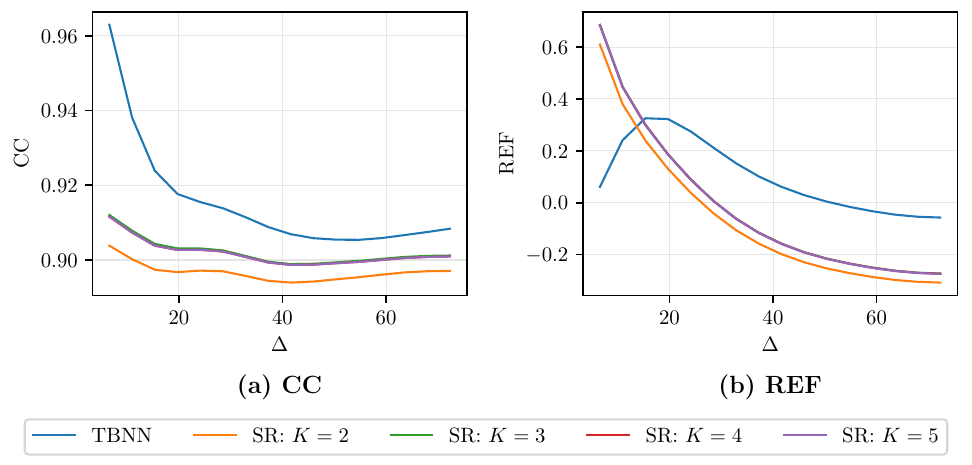}}% Images in 100% size
  \caption{(a) Correlation coefficient (CC) and (b) relative error in mean energy flux (REF) versus filter width ($\Delta$) for models with varying degrees of maximum input order ($K$). Note SR models of maximum order $K=3, 4,$ and $5$ are overlapping.}
    \label{fig:result - order}
\end{figure}

Figure \ref{fig:result - order} shows the variation of (a) CC and (b) REF with $\Delta$ for sparse regression (SR) models trained with $\alpha=1e-4$, $\eta=1/2$, and $\beta=0$. The retained predictors and coefficients for each model are given in \ref{sec:appendix -- model expressions}. For all models, CC is highest at small $\Delta$s and remains largely stable for $\Delta \geq 20$. Across all filter widths, the SR models exhibit slightly lower CC than the TBNN. The REF indicates that the TBNN model is over-dissipative at small $\Delta$ but approaches zero as $\Delta$ increases. The SR models show a similar trend, transitioning from over-dissipative at small $\Delta$s, to near-zero REF around $\Delta \approx 30$, then slightly under-dissipative at larger $\Delta$s. The SR models exhibit negligible sensitivity in either metric to increasing $K$ beyond $3$.

\vspace{3pt}\noindent{\underline{Key Takeaways}}

The TBNN model --- capable of representing very high degrees of nonlinear interactions --- outperforms the SR models in $\textit{a priori}$ tests. Nevertheless, repeatedly evaluating neural networks, even shallow ones, at every discrete point across thousands of iterations during a CFD simulation incurs a significant computational expense. The SR models, in contrast, are polynomial-based with lower, well-defined model orders, yet they still maintain competitive on the evaluated metrics. Although increasing the predictor set order enlarges the candidate tensor library and allows more complex basis representations, no improvement in $\textit{a priori}$ performance is observed beyond $K=3$. This demonstrates that carefully selected lower-order predictor sets can capture key SGS turbulence physics while significantly reducing computational cost compared with higher-order models like TBNNs. As such, we fix $K=3$ for all SR models used in subsequent analyses.

%% ---------------------------
%% A Posteriori Results: Penalty Parameters
%% ---------------------------
\subsection{\textit{A Posteriori} Sensitivity Analysis: Training-Objective Penalty Parameters} \label{sec:penalty parameters}

For data-driven SGS models, the choice of penalty terms in the training objective can have a significant impact on the resulting closure's behavior when deployed in an LES. In this section, we explore two hyperparameters in the sparse regression framework's objective function (Equation~\eqref{eqn:custom objective function}): the regularization penalty, $\alpha$, and the dissipation penalty, $\beta$. The parameter $\alpha$ sets the strength of regularization applied to the model coefficients, balancing $L_1$ sparsity promotion with $L_2$ smoothness. By controlling this trade-off, $\alpha$ influences the model sparsity, efficiency, interpretability, and sensitivity. The parameter $\beta$ weights a dissipation-based constraint, penalizing discrepancies between the modeled and exact SGS dissipation during training. By scaling the relative importance of the dissipation constraint in the objective, $\beta$ influences the extent to which the learned closure reproduces the reference energy transfer between resolved and unresolved scales, impacting physical consistency and numerical stability. Understanding how each of these parameters influence model behavior is critical for striking an appropriate balance between accuracy, efficiency, stability, and robustness in LES deployments.

\begin{table}[t]
    \centering
    \NiceMatrixOptions{custom-line={command = DoubleRule , multiplicity = 2 , sep-color = white}}
    % \NiceMatrixOptions{hvlines-except-borders}
    \begin{NiceTabular}{c|c|c|c|c|c|c|c}
    \DoubleRule
                & \Block{1-1}{\textbf{Reynolds}\\\textbf{Number}}
                & \Block{1-1}{\textbf{Grid}\\\textbf{Resolution}}
                & \Block{1-1}{\textbf{Domain}\\\textbf{Size}}
                & $\Delta x^+$
                & $\Delta y_1^+$
                & $\Delta y_c^+$
                & $\Delta z^+$ \\ \hline
            \Block{3-1}{\textbf{Forced HIT}}
                & \Block{3-1}{$Re_\lambda = \infty$}
                & $32^3$ 
                & \Block{3-1}{$(2\pi)^3$} 
                & \Block{3-1}{-} 
                & \Block{3-1}{-}
                & \Block{3-1}{-}
                & \Block{3-1}{-} \\ \cline{3}
            & & $64^3$ & \\ \cline{3}
            & & $128^3$ & \\ \hline
            \Block{2-1}{\textbf{Taylor-Green}\\\textbf{Vortex Flow}}
                & \Block{2-1}{$Re = 1,\!600$}
                & $64^3$
                & \Block{2-1}{$(2\pi)^3$}
                & \Block{2-1}{-} 
                & \Block{2-1}{-}
                & \Block{2-1}{-}
                & \Block{2-1}{-} \\ \cline{3}
            & & $128^3$ & \\ \hline
            \Block{2-1}{\textbf{Channel}\\\textbf{Flow}}
                & \Block{2-1}{$Re_\tau = 590$}
                & $48 \! \times \! 111 \! \times \! 48$ 
                & \Block{2-1}{$2\pi\delta \! \times \! 2\delta \! \times \! \pi\delta$ 
                % \\ where $delta=$ 1 is \\ the channel \\ half-height
                }
                & 78
                & 1
                & 45
                & 39 \\ \cline{3} \cline{5-8}
            & & $64 \! \times \! 133 \! \times \! 64$ &
                & 58
                & 1
                & 36
                & 29 \\ \hline \hline
    \end{NiceTabular}
    \caption{\textit{A posteriori} flow configurations used for training-objective parameter sensitivity analyses.}
    \label{tab:posteriori tests}
\end{table}

To evaluate these effects in practice, the $\alpha$ and $\beta$ sensitivity studies are performed \textit{a posteriori} using PHASTA, an SUPG/PSPG/grad-div-stabilized finite element-based CFD solver \citep{Whiting01}, to solve the advective form of the incompressible filtered Navier-Stokes equations. We adopt the stabilization matrix formulation detailed in \cite{Tejada-Martinez05} and the stabilization parameters reported in \cite{Prakash22}. Piecewise tri-linear polynomial basis functions are used for the spatial discretization of hexahedral grid elements, and the generalized-$\alpha$ method \citep{Jansen00} is employed for temporal discretization. The local anisotropy tensor $\boldsymbol{A}(\boldsymbol{x})$ is defined from the local geometry of the underlying computational grid according to the procedure outlined in \ref{sec:appendix -- aniso local}. \text{A posteriori} tests in this section focus on three flow configurations: forced HIT, Taylor-Green vortex flow, and turbulent channel flow. Table~\ref{tab:posteriori tests} summarizes the corresponding flow parameters, computational domains, and mesh resolutions. In each of these simulations, the filter is equivalent to the local grid size and the maximum CFL number is $0.5$.

\vspace{3pt}\noindent{\underline{Homogeneous Isotropic Turbulence at $Re_\lambda = \infty$}}

We test the models in \textit{a posteriori} on forced HIT --- consistent with the flow physics used for training --- to assess generalization to an effectively infinite Reynolds number, $Re_\lambda = \infty$, obtained by setting the viscosity to the nominal value of $1\!\times\!10^{-12}$. The simulations are initialized with interpolated instantaneous turbulent velocity and pressure fields from the $Re_\lambda = 418$ simulation used for training \citep{Li08, Perlman07}. Periodic boundary conditions are enforced in all directions, and turbulence is sustained through a low-wavenumber energy forcing following \cite{Prakash22}. In the absence of DNS data at this Reynolds number, the statistically-converged energy spectra are compared to the Kolmogorov energy spectrum:

\begin{equation}
    E(\kappa) = C\epsilon^{2/3}\kappa^{-5/3},
    \label{eqn:kolmogorov}
\end{equation}

\noindent where $\kappa$ is the wavenumber, $C=1.6$ is an empirically-determined constant \cite{Pope00}, and $\epsilon$ is the dissipation rate, which, for a statistically stationary equilibrium state, is equal to the power input of the forcing. To review the LES results of classical SGS models for this flow case, the reader is referred to \cite{Prakash22}.

\vspace{3pt}\noindent{\underline{Taylor-Green Vortex Flow at $Re=1,600$}}

To assess \textit{a posteriori} performance on unsteady, non-equilibrium turbulence, we consider the three-dimensional Taylor–Green vortex, a flow configuration featuring dynamics not present in the training dataset. The simulation begins in a laminar regime, transitions to turbulence at a sufficiently high Reynolds number, then decays after reaching a fully developed turbulent state. The velocity field is initialized as:

\begin{equation}
    \mathbf{u}=
        \left[\begin{array}{c}
              \sin \left(x_1\right) \cos \left(x_2\right) \cos \left(x_3\right) \\
              -\cos \left(x_1\right) \sin \left(x_2\right) \cos \left(x_3\right) \\
              0 \end{array}\right].
    \label{eqn:TGV IC}
\end{equation}

\noindent Periodic boundary conditions are applied on all faces of the cubic domain. Model performance is assessed by tracing the temporal evolution of the resolved dissipation:

\begin{equation}
    \epsilon(t)=
        2 v \frac{1}{\Omega} \int_{\Omega} \frac{\boldsymbol{\omega}(t) 
            \cdot \boldsymbol{\omega}(t)}{2} d \Omega ,
    \label{eqn:TGV diss}
\end{equation}

\noindent where $\boldsymbol{\omega}$ is the vorticity. The LES results are compared to DNS data from \cite{Shoraka17}. For LES results obtained with classical SGS models for this flow case, see \cite{Prakash22}.

\vspace{3pt}\noindent{\underline{Turbulent Channel Flow at $Re_\tau=590$}}

Wall-resolved LES of turbulent channel flow are performed to evaluate model performance in a wall-bounded flow with anisotropic grid resolutions not represented in the training dataset. The flow is initialized with a log-law velocity profile perturbed by random Gaussian noise. Periodic boundary conditions are applied in the streamwise and spanwise directions, while no-slip boundary conditions are applied at $y=0$ and $y=2\delta$ (where $\delta$ is the channel half-height). A constant mass flux forcing, corresponding to a bulk Reynolds number $Re_b$ of $10,975$, sustains the flow. The nondimensional grid spacing in each Cartesian direction is provided in Table~\ref{tab:posteriori tests}, where $\Delta x^+$ is the streamwise grid spacing, $\Delta y_1^+$ is the wall-normal grid spacing for the first off-wall element, $\Delta y_c^+$ is the wall-normal grid spacing at the channel centerline, and $\Delta z^+$ is the spanwise grid spacing. Once statistical stationarity is reached, LES velocity profiles are averaged and compared to the DNS results from \cite{Moser99}. Classical SGS model LES results for this flow case can be found in \cite{Prakash24}.

%% ---------------------------
%% A Posteriori Results: Alpha
%% ---------------------------

\begin{figure}[b]
\centerline{\includegraphics[width=0.99\textwidth]{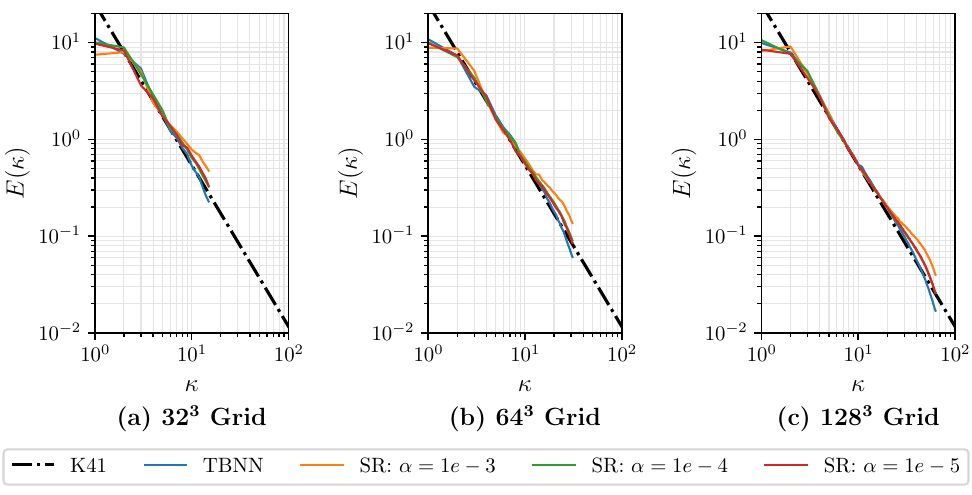}}% Images in 100% size
  \caption{Three-dimensional energy spectra of forced HIT at $Re_\lambda = \infty$ on (a) $32^3$, (b) $64^3$, and (c) $128^3$ isotropic grids for models with varying Lasso penalty parameters ($\alpha$). Note that the SR spectra with $\alpha=1e-4$ and $1e-5$ largely coincide at higher wavenumbers.}
    \label{fig:result - HIT alpha}
\end{figure}

\subsubsection{Effect of Regularization Penalty ($\alpha$)} \label{sec:alpha}

We explore the effect of varying the regularization penalty parameter $\alpha$. Throughout this study, the remaining SR objective function hyperparameters are held fixed ($K=3$, $\eta=1/2$, and $\beta=0$). \ref{sec:appendix -- model expressions} presents the downselected predictors and coefficients for each SR model analyzed here. The objective of this analysis is not to determine an optimal value of $\alpha$, as the appropriate balance between regularization and performance is problem-dependent and varies with flow regime, modeling goals, and computational constraints. Instead, we investigate how variations in $\alpha$ influence a model's ability to capture key turbulence physics across a range of canonical flow cases. 

%% ---------------------------
%% A Posteriori Results: Alpha -- HIT
%% ---------------------------
\vspace{3pt}\noindent{\underline{Homogeneous Isotropic Turbulence at $Re_\lambda = \infty$}}

\begin{figure}[b]
\centerline{\includegraphics[width=0.99\textwidth]{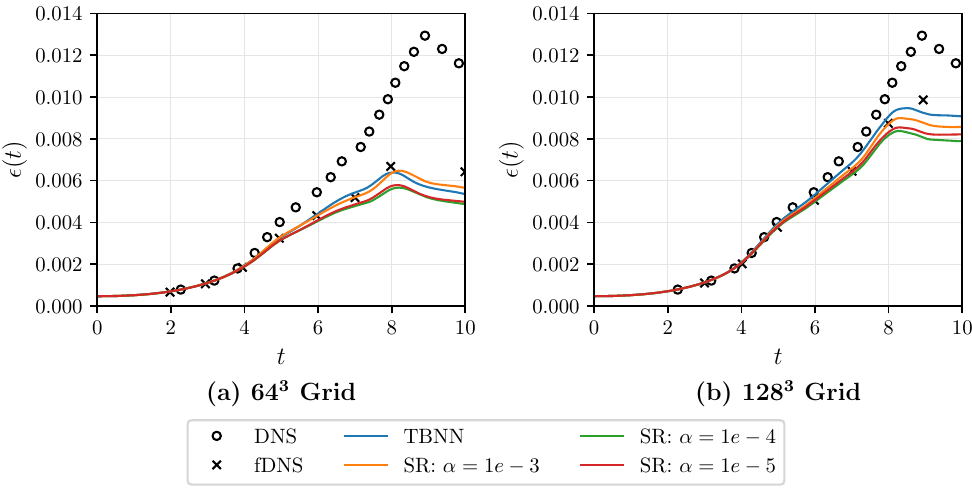}}% Images in 100% size
  \caption{Temporal evolution of the resolved dissipation for Taylor-Green vortex flow at $Re = 1,600$ on (a) $64^3$ and (b) $128^3$ grid resolutions for models with varying Lasso penalty parameters ($\alpha$).}
    \label{fig:result - TGV eps alpha}
\end{figure}

The $\textit{a posteriori}$ sensitivity to $\alpha$ is first assessed on forced HIT at a higher Reynolds number than used for training. Figure~\ref{fig:result - HIT alpha} shows the three-dimensional energy spectra on (a) $32^3$, (b) $64^3$, and (c) $128^3$ grids. Across all resolutions, the models recover the expected $\kappa^{-5/3}$ scaling at intermediate wavenumbers. Differences emerge at larger wavenumbers: the TBNN model maintains good agreement with the reference spectrum, the $\alpha=1e-3$ SR model exhibits appreciable energy pile-up, and the $\alpha=1e-4$ and $\alpha=1e-5$ SR models exhibit nearly identical performance with only slight energy pile-up. This trend indicates that excessive regularization degrades the representation of the impact of small-scale dynamics and is more likely to underpredict the SGS dissipation. As $\alpha$ is reduced, fine-scale turbulent structures are more accurately captured and performance approaches that of the TBNN. Reducing $\alpha$ beyond $1e-4$ yields diminishing returns despite increased model complexity, suggesting the inclusion of redundant or weakly correlated terms as regularization is weakened. 

%% ---------------------------
%% A Posteriori Results: Alpha -- TGV
%% ---------------------------
\vspace{3pt}\noindent{\underline{Taylor-Green Vortex Flow at $Re=1,600$}}

We next analyze the effect of $\alpha$ variation for the three-dimensional Taylor-Green vortex flow at $Re=1,600$. Figure~\ref{fig:result - TGV eps alpha} shows the temporal evolution of the resolved dissipation on the (a) $64^3$ and (b) $128^3$ grids. On both meshes, the TBNN and SR model with $\alpha=1e-3$ remain in good agreement with the fDNS results through $t=8$, after which they underpredict dissipation as the coherent structures break down and the flow begins to decay. Curiously, SR models with weaker regularization (i.e. increased model density) do not tend toward the predictions of the TBNN. Rather, the $\alpha=1e-4$ and $\alpha=1e-5$ SR models deviate further from them as the flow evolves beyond the inviscid phase. We posit that this behavior arises from differences in the underlying functional representation of the two model classes. Similar behavior is observed in the energy spectra at $t=9$ in Figure~\ref{fig:result - TGV spectra alpha} of \ref{sec:appendix -- alpha}. 

%% ---------------------------
%% A Posteriori Results: Alpha -- Channel
%% ---------------------------
\vspace{3pt}\noindent{\underline{Turbulent Channel Flow at $Re_\tau=590$}}

\begin{figure}[b]
\centerline{\includegraphics[width=0.99\textwidth]{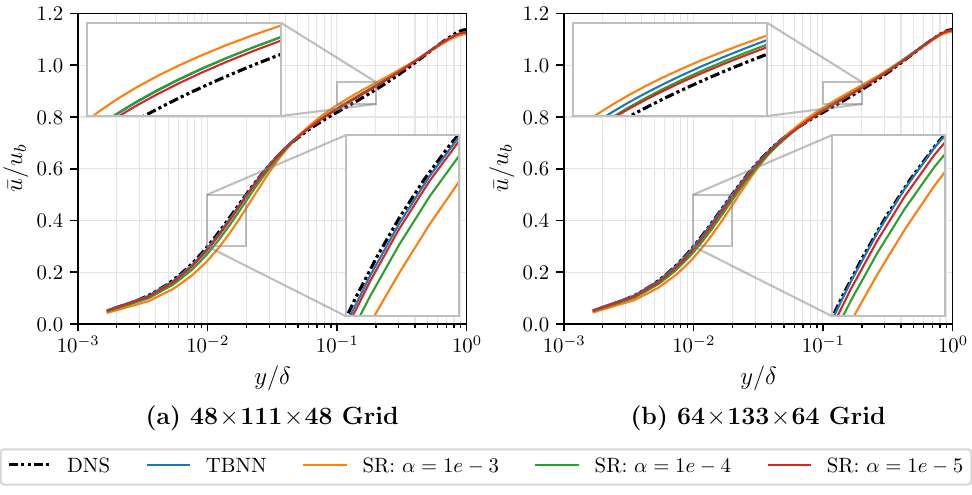}}% Images in 100% size
  \caption{Streamwise velocity profiles for turbulent channel flow at $Re_\tau = 590$ on the (a) $48\times111\times48$ and (b) $64\times133\times64$ grid resolutions for models with varying Lasso penalty parameters ($\alpha$). Note that the SR model with $\alpha=1e-5$ and TBNN profiles overlap in the outer-region on the coarser grid.}
    \label{fig:result - Channel u alpha}
\end{figure}

Lastly, we examine the impact of $\alpha$ variation on SR model performance in turbulent channel flow at $Re_\tau=590$. Figure~\ref{fig:result - Channel u alpha} shows the streamwise velocity profiles for the (a) coarse and (b) medium mesh resolutions. The TBNN velocity profiles coincide well with the DNS results near the wall but deviate further away, while the $\alpha=1e-5$ SR model closely reproduces the DNS profiles in all regions of the flow, often recovering or exceeding TBNN performance. Increasing regularization amplifies underprediction of the near-wall velocities and overprediction further from the wall, a trend consistent across both meshes. The least regularized model avoids these discrepancies, likely given its increased representational capacity, leading to enhanced accuracy in predictions of the steep gradients and anisotropic features of this wall-bounded flow. Corresponding Reynolds stress components are provided in Figure~\ref{fig:result - Channel RS alpha} in \ref{sec:appendix -- alpha} for reference.

%% ---------------------------
%% A Posteriori Results: Alpha -- Summary
%% ---------------------------
\vspace{3pt}\noindent{\underline{Key Takeaways}}

Each canonical flow case considered here exposes a different consequence of model regularization, revealing how it shapes model expressiveness, fidelity, and generalization. In HIT, a flow aligned with the training data, excessive regularization leads to underpredicted SGS dissipation and a pile-up of high-wavenumber energy. Reducing model regularization improves fidelity up to a point, beyond which additional model complexity offers negligible gains in accuracy, indicating a possible saturation of the physically relevant feature space for this flow. In contrast, for the transitional Taylor-Green Vortex, improved agreement with the reference results is observed for the more strongly regularized SR models. In the turbulent channel flow, however, reduced regularization is beneficial; the increased representational flexibility allows the model to better predict the strong near-wall gradients and anisotropic structures absent from the training data. Taken together, these results demonstrate that $\alpha$ highlights the role of regularization in mediating the balance between model parsimony, predictive performance, and robustness, with its effects varying across flow regimes. 

This sensitivity analysis suggests that, under suitable regularization, SR models can achieve predictive capabilities comparable to the TBNN and, in some regimes, even exceed them. This is notable, as SR models incur a lower computational cost than TBNNs, with regularization further modulating cost through its control of model sparsity, enabling efficient yet physically informed predictions.

%% ---------------------------
%% A Posteriori Results: Beta
%% ---------------------------
\subsubsection{Effect of Dissipation Penalty ($\beta$)} \label{sec:beta}

\begin{figure}[t]
\centerline{\includegraphics[width=0.99\textwidth]{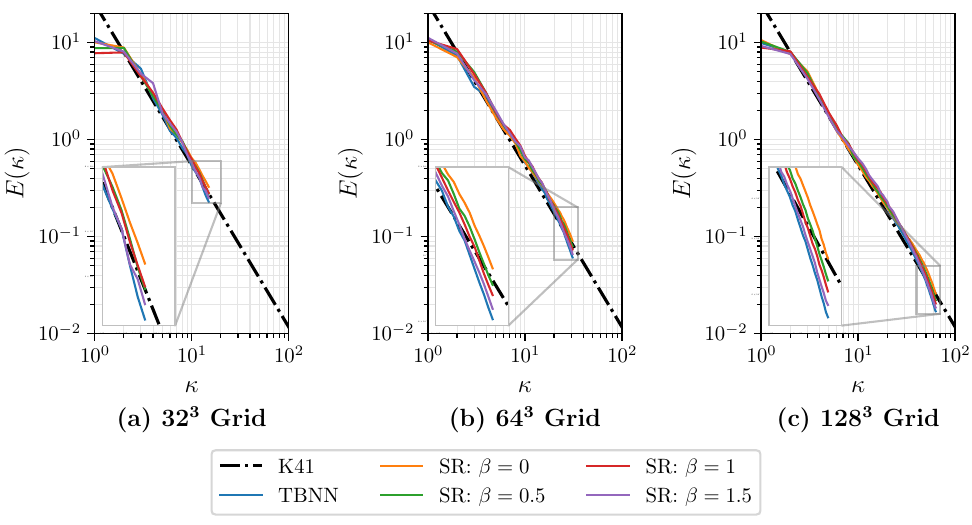}}% Images in 100% size
  \caption{Three-dimensional energy spectra of forced HIT at $Re_\lambda = \infty$ on (a) $32^3$, (b) $64^3$, and (c) $128^3$ isotropic grids for models with varying penalty parameters on the SGS dissipation correction ($\beta$).}
    \label{fig:result - HIT beta}
\end{figure}

We examine the impact of varying the SGS dissipation penalty parameter $\beta$. All SR models considered up to this point have been generated with $\beta=0$, corresponding to the elastic net objective defined in Equation~\eqref{eqn:elastic net}. In the results that follow, $\beta$ is systematically increased such that the training objective penalizes deviations between the modeled SGS dissipation and that predicted by the fDNS data. Throughout this analysis, the remaining SR hyperparameters are held constant ($K=3$, $\eta=1/2$, and $\alpha=1e-4$), with $\alpha$ chosen based on Section~\ref{sec:alpha} to reflect an intermediate level of regularization, providing a balanced representation of key turbulence physics while still allowing room for further refinement. The predictors and coefficients for each SR model studied here are provided in \ref{sec:appendix -- model expressions}. As in the sensitivity study of $\alpha$, the aim is not to identify an optimal $\beta$ value, but rather to examine the influence of dissipation-aware training on model behavior.

%% ---------------------------
%% A Posteriori Results: Beta -- HIT
%% ---------------------------
\vspace{3pt}\noindent{\underline{Homogeneous Isotropic Turbulence at $Re_\lambda = \infty$}}

As before, we first assess model performance on the same flow regime used for training: forced HIT. Figure~\ref{fig:result - HIT beta} presents the three-dimensional energy spectra for SR models with varying $\beta$ across three grid resolutions. At low wavenumbers, all models closely follow K41 scaling, with enhanced alignment on finer grids. At intermediate wavenumbers, the models begin to overestimate energy, particularly those with higher $\beta$ values. At high wavenumbers, where dissipation dominates, low values of $\beta$ lead to energy pile-up while high $\beta$-valued models become over-dissipative. These trends --- consistent across all grid resolutions --- likely stems from the stronger emphasis on unresolved-scale dissipation imposed by larger $\beta$ values during training. This can over-constrain the dynamics of the smallest resolved scales, disrupting energy transfer through the cascade and causing accumulation at intermediate wavenumbers.

%% ---------------------------
%% A Posteriori Results: Beta -- TGV
%% ---------------------------
\vspace{3pt}\noindent{\underline{Taylor-Green Vortex Flow at $Re=1,600$}}

\begin{figure}[t]
\centerline{\includegraphics[width=0.99\textwidth]{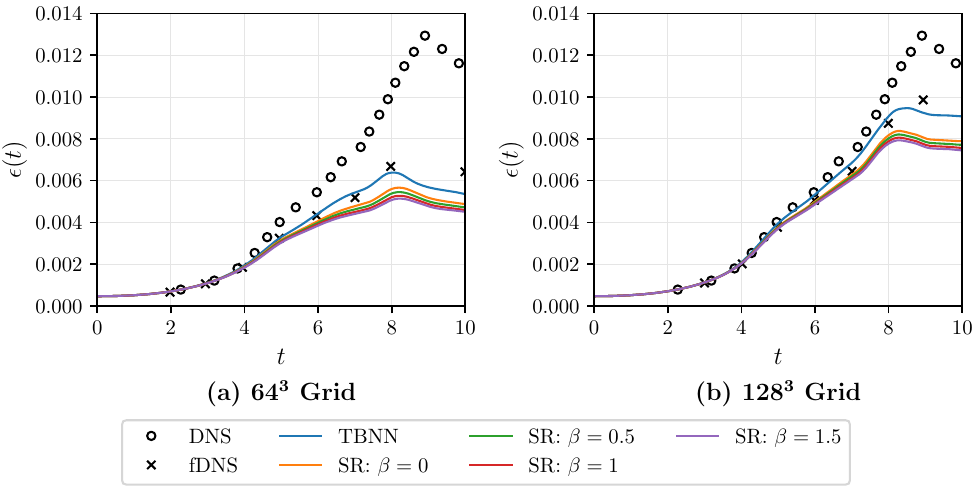}}% Images in 100% size
  \caption{Temporal evolution of the resolved dissipation for Taylor-Green vortex flow at $Re = 1,600$ on (a) $64^3$ and (b) $128^3$ grid resolutions for models with varying penalty parameters on the SGS dissipation correction ($\beta$).}
    \label{fig:result - TGV eps beta}
\end{figure}

We now consider the Taylor-Green vortex at $Re=1,600$, examining the effect of varying $\beta$ on the temporal evolution of the resolved dissipation in Figure~\ref{fig:result - TGV eps beta}. As the flow progresses into the dissipation-dominated decay phase, the SR models increasingly underpredict the resolved dissipation, with discrepancy growing as $\beta$ increases. This trend indicates that larger $\beta$ values reduce the amount of dissipation sustained at resolved scales throughout the decay. This behavior can be interpreted in the context of the energy cascade: higher $\beta$ values place greater emphasis on unresolved-scale dissipation in the training objective, yielding closures that prematurely remove energy from the smallest resolved scales and effectively truncate the intermediate scales of the energy cascade. The resulting energy deficit at resolved scales leads to the observed underprediction of dissipation. This imbalance in the dissipation process is particularly pronounced for Taylor–Green vortex flow, where accurate dissipation across all scales is crucial. Consistent behavior is observed in the corresponding energy spectra, recorded at a time near maximum dissipation, in Figure~\ref{fig:result - TGV spectra beta} provided in \ref{sec:appendix -- beta}.

%% ---------------------------
%% A Posteriori Results: Beta -- Channel
%% ---------------------------
\vspace{3pt}\noindent{\underline{Turbulent Channel Flow at $Re_\tau=590$}}

\begin{figure}[t]
\centerline{\includegraphics[width=0.99\textwidth]{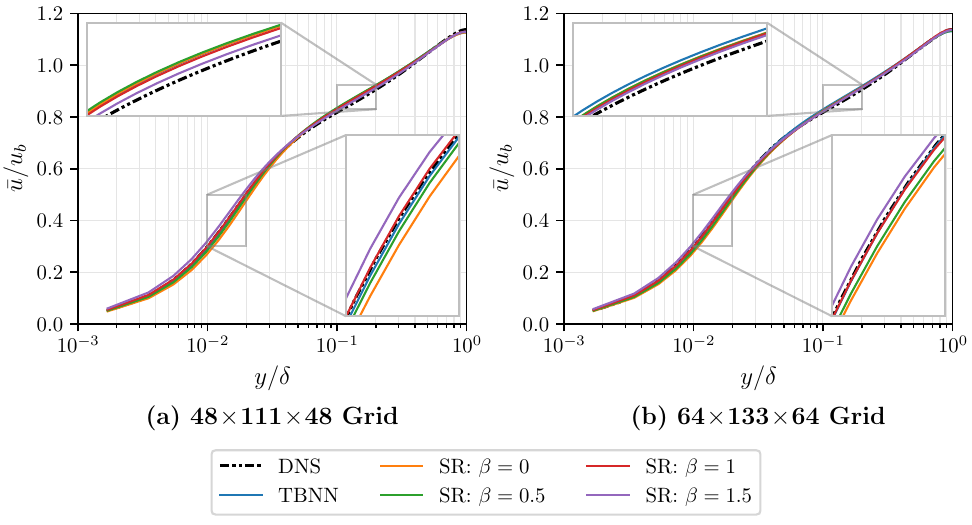}}% Images in 100% size
  \caption{Streamwise velocity profiles for turbulent channel flow at $Re_\tau = 590$ on the (a) $48\times111\times48$ and (b) $64\times133\times64$ grid resolutions for models with varying penalty parameters on the SGS dissipation correction ($\beta$). Note that, on the finer mesh, the SR model with $\beta=1$ and TBNN profiles overlap in the near-wall region.}
    \label{fig:result - Channel u beta}
\end{figure}

Next, we examine $\beta$'s effect on the streamwise velocity profiles in turbulent channel flow, shown in Figure~\ref{fig:result - Channel u beta}. Increasing $\beta$ generally improves alignment with DNS velocity profiles on both grid resolutions. However, excessive SGS dissipation correction ($\beta=1.5$) leads to overprediction of the near-wall streamwise velocities. SR models with intermediate $\beta$ values perform well across the channel. In particular, the $\beta=1$ SR model achieves accuracy comparable to the TBNN in the near-wall region and outperforms in the outer-region. These trends become more pronounced as the grid resolution is refined. In \ref{sec:appendix -- beta}, corresponding Reynolds stress component profiles are shown in Figure~\ref{fig:result - Channel RS beta}.

%% ---------------------------
%% A Posteriori Results: Beta -- Summary
%% ---------------------------
\vspace{3pt}\noindent{\underline{Key Takeaways}}

The dissipation penalty alters model behavior by reweighting the enforcement of SGS dissipation correction during training, redistributing energy across scales. These effects are distinct among the canonical flows considered. In forced HIT, increasing the dissipation penalty reduces high-wavenumber energy pile-up but disrupts the energy transfer through the inertial range, producing excess energy at intermediate scales. This reflects how stronger weighting of subgrid dissipation during model training can over-constrain the cascade, redistributing energy transfer rather than producing consistent improvements across the spectrum. In the Taylor-Green vortex flow, stronger dissipation penalization exacerbates the underprediction of resolved dissipation over time. This behavior is consistent with premature energy removal at unresolved scales, limiting intermediate-scale energy transfer and diminishing dissipation at resolved scales. In contrast, the turbulent channel flow generally exhibits improved velocity predictions with moderately large dissipation penalties, while excessive dissipation enforcement leads to overprediction of near-wall velocities. Overall, these results indicate that $\beta$ influences the distribution of dissipation across resolved and unresolved scales, mediating the balance between functional and structural performance, and manifesting differently across the flow cases considered.

This analysis demonstrates that dissipation-aware training objectives can directly shape the energetic behavior of SR models, where the effect is governed by the characteristics of the training data. By training on forced HIT --- a flow in which statistical stationarity is maintained through artificial energy injection in an idealized homogeneous setting --- the dissipative behaviors available to inform this modeling constraint are inherently limited. These results suggest that training over flows with more realistic, non-equilibrium cascade dynamics could further improve model performance without significantly increasing model cost or complexity.

%% ---------------------------
%% A Posteriori Results: Periodic Hill
%% ---------------------------
\subsection{\textit{A Posteriori} Predictive Performance Benchmark: Separated Flow} \label{sec:periodic hill}

To evaluate the generalization capability of the proposed SGS modeling framework beyond the canonical flows considered in the sensitivity analysis, we next consider the periodic hill benchmark. This flow regime introduces geometry-induced separation, recirculation, and reattachment, providing a test of model behavior under flow conditions that differ fundamentally from those represented in the training data. Because the focus in this section is on generalization as opposed to cross-model comparison, results are presented for only a single SR model (with $K=3$, $\alpha=1e-4$, and $\beta=1$) selected from the sensitivity analysis as a representative configuration. The performance of this SR model is compared against the DNS results reported in \cite{Breuer09}, the TBNN model, and the dynamic Smagorinsky (dynSmag) model (where the scalar characteristic filter width is taken as the geometric mean of the local grid spacing) to provide context for its predictive performance. 

\begin{wrapfigure}{l}{0.475\textwidth}
    \includegraphics[width=0.475\textwidth]{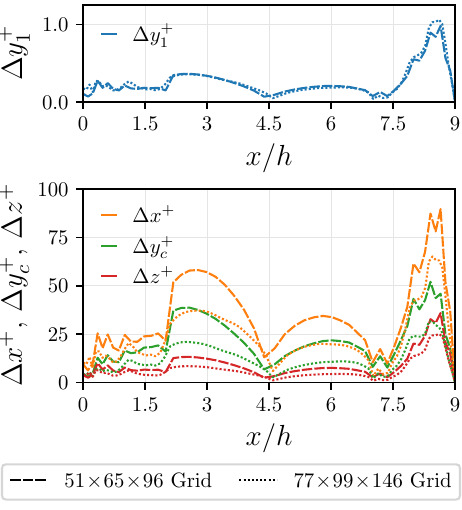}
    \caption{Distribution of nondimensional grid spacing along the lower wall ($\Delta x^+$, $\Delta y_1^+$, $\Delta z^+$) and the channel centerline ($\Delta y_c^+$) for the periodic hill cases.}
        \label{fig:hill plus}
\end{wrapfigure}

The periodic hill is simulated using wall-resolved LES within the same numerical framework employed for the \textit{a posteriori} sensitivity analyses, PHASTA. The dimensions of the channel --- $9 \!\times\! 3.035 \!\times\! 4.5$ --- follow the convention set forth by \cite{Mellen00}, and the shape of the hill --- with height $h=1$  --- is taken from \cite{Almeida93}. The Reynolds number $Re_h=u_bh/\nu$ of $5,600$ is based on the hill height and the bulk velocity above the crest of the first hill. Periodic boundary conditions are applied in the streamwise and spanwise directions to emulate an infinite sequence of hills. No-slip boundary conditions are applied at the top and bottom walls. A constant mass forcing is used to sustain the flow. Simulations are conducted across two grid resolutions --- on a coarser mesh of $51\times65\times96$ elements and a finer mesh of $77\times99\times146$ elements. The nondimensional grid spacing for each grid resolution is plotted in Figure~\ref{fig:hill plus}.

The predictive performance of the models is evaluated based on their ability to reproduce key flow features of the periodic hill, including separation and reattachment locations, shear stress and pressure distributions along the bottom wall, and velocity profiles at representative streamwise stations (Reynolds stress profiles are provided in \ref{sec:appendix -- periodic hill} for reference). The predicted separation ($x_{sep}/h$) and reattachment ($x_{att}/h$) locations are summarized in Table~\ref{tab:result - hill points}. On the coarser grid, the data-driven models predict an earlier onset of separation compared to the DNS reference, whereas the dynSmag model yields delayed separation. With increased resolution, all models shift the separation point downstream relative to DNS. Across both meshes, the SR model provides the closest agreement with DNS for separation location, while the dynSmag model shows the largest deviation. Reattachment is predicted prematurely by all models at both resolutions, with the TBNN aligning the most closely with DNS, while the dynSmag again demonstrates the greatest discrepancy.

\begin{figure}[t]
\centerline{\includegraphics[width=0.99\textwidth]{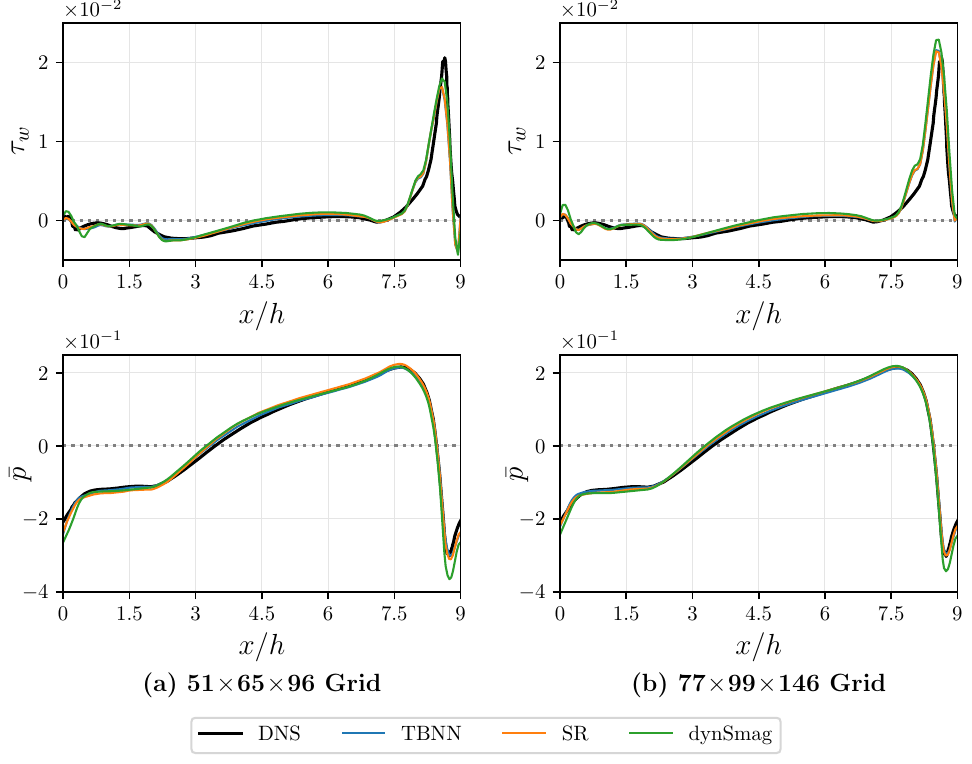}}% Images in 100% size
  \caption{Averaged wall shear stress and pressure distributions along the lower wall of the periodic hill at $Re_h=5,600$ on (a) the coarser and (b) the finer grid resolutions. Note that the SR and TBNN models largely coincide for both metrics.}
    \label{fig:result - Hill wall}
\end{figure}

{\parfillskip=0pt
The streamwise distribution of wall shear stress, $\tau_w$, along the lower wall of the periodic hill is shown in Figure~\ref{fig:result - Hill wall} for both grid resolutions. The TBNN and SR models coincide closely across most of the domain. Consistent with the separation and reattachment locations reported in Table~\ref{tab:result - hill points}, separation is marked by a change in sign of $\tau_w$ from positive to negative, while reattachment is indicated by the return to positive values. Downstream of reattachment, near the foot of the hill at $x/h\approx7.2$, all models predict a small secondary recirculation zone 
\par}

\begin{wraptable}{r}{8.65cm}
    \centering
    \NiceMatrixOptions{custom-line={command = DoubleRule , multiplicity = 2 , sep-color = white}}
    \begin{NiceTabular}{c|c|c|c}
    \DoubleRule
    & \textbf{Model} & $\bm{x_{sep}/h}$ & $\bm{x_{att}/h}$ \\ \cline{2-4}
    & DNS     & 0.18 & 5.1 \\ \hline
    \Block{3-1}{$\mathbf{51\times65\times96}$\\\textbf{Grid}}
    & TBNN    & 0.14 & 4.7 \\ \cline{2-4}
    & SR      & 0.15 & 4.5 \\ \cline{2-4}
    & dynSmag & 0.28 & 4.3 \\ \hline
    \Block{3-1}{$\mathbf{77\times99\times146}$\\\textbf{Grid}}
    &TBNN    & 0.22 & 4.8 \\ \cline{2-4}
    &SR      & 0.21 & 4.6 \\ \cline{2-4}
    &dynSmag & 0.25 & 4.5 \\
    \hline\hline
    \end{NiceTabular}
    \caption{Separation ($x_{sep}/h$) and reattachment ($x_{att}/h$) locations for the periodic hill at $Re_h=5,600$.}
    \label{tab:result - hill points}
\end{wraptable}

\noindent before $\tau_w$ rises to a maximum near the hill crest. On the coarser grid, all models underpredict the peak stress magnitude, with the dynSmag showing the closest agreement with DNS; on the finer grid, all overpredict, with the SR model providing the closest alignment. Across both meshes and all models, peak stresses are predicted slightly upstream of the reference. The corresponding averaged wall pressure distribution, $\bar{p}$, is also shown in Figure~\ref{fig:result - Hill wall}. All models reproduce the post-separation pressure plateau reasonably well, with the TBNN showing the best agreement with DNS. As pressure increases downstream, each of the models overestimate its magnitude through reattachment near $x/h\approx5.1$ --- with deviation more pronounced on the coarser mesh. On the finer grid, the magnitude and location of the pressure maximum near $x/h\approx7.8$ is well reproduced by all models, as is the subsequent sharp pressure drop. On the coarser grid, the SR model slightly exaggerates the peak pressure magnitude, while the dynSmag model predicts the maximum too far upstream, resulting in a premature decline relative to the reference data. The dynSmag model also overshoots the pressure minimum on both mesh resolutions, whereas the data-driven models remain closer to the reference, with the SR model exhibiting a modest overprediction on the coarser grid.

\begin{figure}
\centerline{\includegraphics[width=0.99\textwidth]{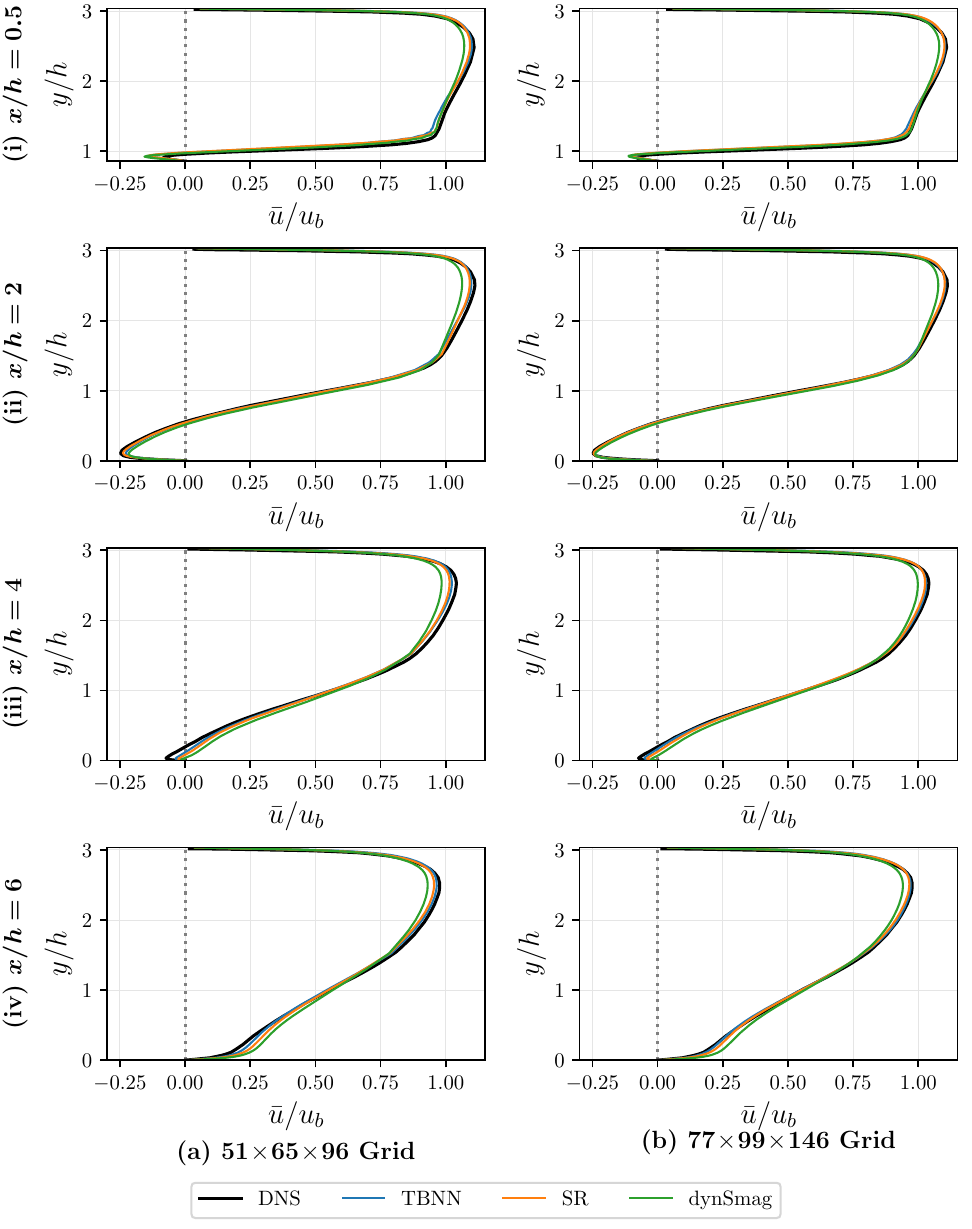}}% Images in 100% size
  \caption{Streamwise velocity profiles for the periodic hill at $Re_h=5,600$ on (a) the coarser and (b) the finer grid resolutions at (\romannumeral1\relax) $x/h=0.5$, (\romannumeral2\relax) $x/h=2$, (\romannumeral3\relax) $x/h=4$, and (\romannumeral4\relax) $x/h=6$.}
    \label{fig:result - Hill x-vel}
\end{figure}

Figure~\ref{fig:result - Hill x-vel} shows averaged streamwise velocity profiles for the (a) coarser and (b) finer meshes of the periodic hill. These profiles are plotted at representative streamwise stations: (\romannumeral1\relax) at $x/h=0.5$, along the leeward side of the hill just after the point of separation; (\romannumeral2\relax) at $x/h=2$, at the base of the hill within the primary recirculation region; (\romannumeral3\relax) at $x/h=4$, on the flat channel floor toward the end of the primary recirculation region; and (\romannumeral4\relax) at $x/h=6$, on the flat channel floor after the reattachment point. Within the separation bubble, the models transition from overprediction of the near-wall minimum velocity magnitude at $x/h=0.5$, to close agreement at $x/h=2$, to underprediction at $x/h=4$. These discrepancies coincide with the premature reattachment and shortened recirculation region identified in Figure~\ref{fig:result - Hill wall}(\romannumeral1\relax) and Table~\ref{tab:result - hill points}. The models' early reattachment results in thinner, stronger shear layers downstream at $x/h = 6$, producing steeper velocity gradients relative to the reference data. Alignment to DNS results improves on the finer grid at all stations, where the data-driven models yield reliable predictions in both the near-wall and bulk flow regions, consistently outperforming dynSmag.

\begin{figure}[b]
\centerline{\includegraphics[width=0.99\textwidth]{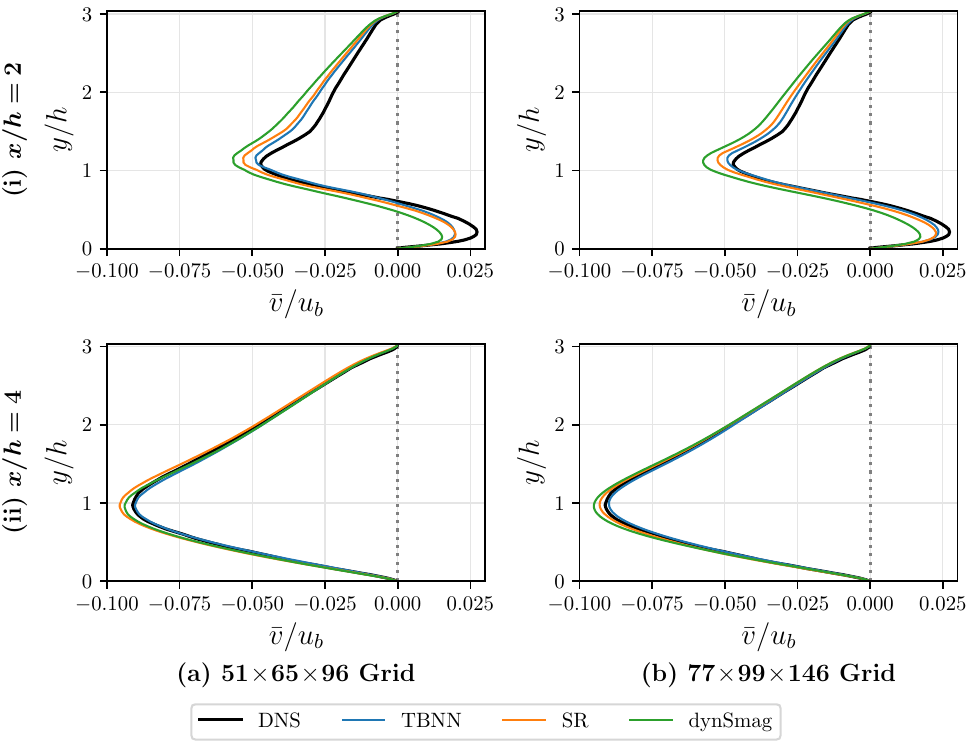}}% Images in 100% size
  \caption{Vertical velocity profiles for the periodic hill at $Re_h=5,600$ on (a) the coarser and (b) the finer grid resolutions at (\romannumeral1\relax) $x/h=2$ and (\romannumeral2\relax) $x/h=4$.}
    \label{fig:result - Hill y-vel}
\end{figure}

The averaged vertical velocity profiles on the (a) coarser and (b) finer meshes for the periodic hill are shown in Figure~\ref{fig:result - Hill y-vel} at (\romannumeral1\relax) $x/h=2$ and (\romannumeral2\relax) $x/h=4$, both within the primary separation bubble. At $x/h=2$, all models underestimate the peak velocity magnitude in the near-wall region while overestimating the minimum velocity magnitude in the outer region, causing profile discrepancies that extend up to the top-wall. At this $x$-location, the TBNN and SR models perform comparably and outperform dynSmag; for all models, deviations from the reference improve with grid refinement. At $x/h=4$, all models align more closely with the DNS results, with the SR and dynSmag models slightly overpredicting the minimum velocity magnitude at each grid resolution, and the TBNN reproducing it closely. 

\vspace{3pt}\noindent{\underline{Key Takeaways}}

The periodic hill benchmark highlights the potential of sparse data-driven turbulence models to capture key physics in flows with complex features such as separation, recirculation, and strong adverse pressure gradients. Across both grid resolutions, the selected SR model provides the closest agreement for the onset of separation, while the TBNN most accurately captures the reattachment location. All models exhibit measurable deviations from DNS predictions of separation and reattachment locations, as these are regions of strong sensitivity, where the response to modeling error is amplified. Wall shear stress and pressure distributions show that the data-driven models nearly coincide through the extent of the domain and reproduce well the principal flow features, with grid refinement improving peak magnitude predictions. Streamwise velocity behavior is well represented by both the SR and TBNN models, with the TBNN exhibiting modestly better predictions in the near-wall region; on the finer mesh, both models achieve strong agreement with DNS in the near-wall and outer regions. Within the central region of the separation bubble, consistent discrepancies arise in the vertical velocity profiles, where the SR model is outperformed by the TBNN. Grid refinement yields modest improvement in this region, but more pronounced alignment emerges downstream as reattachment is approached. 

Despite the absence of features targeted for the periodic hill flow case in the training data, the SR model consistently outperforms the dynSmag model across all evaluated metrics and both grid resolutions. Moreover, it delivers accuracy comparable to the TBNN while retaining a markedly simpler, sparse model form. This benchmark establishes a strong foundation for further exploration and refinement, particularly as this framework is extended to more complex geometries and higher-Reynolds-number regimes.

%% ---------------------------
%% Results: Computational Efficiency
%% ---------------------------
\subsection{Computational Efficiency Benchmark} \label{sec:efficiency}

Predictive accuracy alone does not fully characterize the practical utility of a closure model. Data-driven closures often incur substantially higher costs than classical models, making computational efficiency a critical factor in determining feasibility; even highly accurate models may be impractical if their training and inference overhead is excessive. In this section, we quantify and compare the efficiency of the TBNN model and a representative SR model (with $K=3$, $\alpha=1e-4$, and $\beta=1$, as used in the periodic hill benchmark) during both the training and inference phases. 

The training cost associated with each data-driven modeling approach is assessed in terms of the number of trainable parameters, the wall clock time, and the peak memory usage. Both models are trained on the same dataset (consisting of $1,258,290$ samples) and executed on the same Apple M1 Max workstation with a 32-core integrated GPU. In both cases, GPU acceleration is utilized: the TBNN is implemented in TensorFlow/Keras with a Metal backend, whereas the SR model is implemented in PyTorch with an MPS backend. All training runs are executed in serial to ensure a consistent comparison of algorithmic cost. As shown in Table~\ref{tab:efficiency}, the training procedure for the TBNN is approximately an order of magnitude slower and reaches a peak memory usage over three times that of the SR model. These efficiency differences arise from the training mechanisms. The TBNN relies on backpropagation over $1,500$ epochs through a neural network with $308$ trainable parameters, employing an adaptive learning rate and a batch size of $60,000$. In contrast, the SR model leverages coordinate descent to optimize a much simpler parametric form with only $23$ trainable parameters. By avoiding the overhead of storing and propagating intermediate activations and gradients, the SR model achieves significant reductions in both training wall-clock time and memory usage while retaining comparable predictive capability to the TBNN (as demonstrated in Sections~\ref{sec:alpha}, \ref{sec:beta}, and \ref{sec:periodic hill}). 

\begin{table}[t]
    \centering
    \NiceMatrixOptions{custom-line={command = DoubleRule , multiplicity = 2 , sep-color = white}}
    \begin{NiceTabular}{c|c|c|c|c|c}
    \DoubleRule
        & \Block{1-3}{\textbf{Training}} 
            & & & \Block{1-2}{\textbf{Inference}} \\ \cline{2-6}
        & \Block{1-1}{\textbf{Number of}\\\textbf{Trainable}\\\textbf{Parameters}}
            & \Block{1-1}{\textbf{Wall Clock}\\\textbf{Time}\\\textbf{(Minutes)}}
            & \Block{1-1}{\textbf{Peak}\\\textbf{Memory}\\\textbf{(GB)}}
            & \Block{1-1}{\textbf{Number of}\\\textbf{Retained}\\\textbf{Parameters}}
            & \Block{1-1}{\textbf{Cost per}\\\textbf{Evaluation}\\\textbf{(FLOPs)}} \\ \hline
        \textbf{TBNN}
            & $308$
            & $7.75$
            & $2.47$
            & $308$
            & $1,052$\\ \cline{1-6}
        \textbf{SR}
            & $23$
            & $0.64$
            & $0.75$
            & $14$
            & $437$ \\ 
    \DoubleRule
    \end{NiceTabular}
    \caption{Training and inference costs for the TBNN and a representative SR model (with $K=3$, $\alpha=1e-4$, and $\beta=1$) across key metrics.}
    \label{tab:efficiency}
\end{table}

Evaluation cost is assessed in terms of the number of retained parameters and the associated computational effort per evaluation, quantified in floating-point operations (FLOPs) required for closure evaluation relative to a baseline solve without an SGS model. As shown in Table~\ref{tab:efficiency}, the SR model retains only $14$ nonzero coefficients following training, compared to the full $308$ trainable parameters for the TBNN. This reduction in retained parameters translates into lower computational expense upon inference: the SR model requires less than half the FLOPs of the TBNN for a single evaluation. In a finite element (or finite volume) solver, the FLOPs per evaluation for the data-driven models correspond to the FLOPs per quadrature point per element (or control volume) per nonlinear solve per time step. Consequently, even modest reductions in the size or parameter count of a data-driven model can yield substantial improvements in overall simulation coast, especially in unsteady, high-resolution simulations where stress fields are repeatedly evaluated across thousands (or millions, or even billions) of elements for each time step iteration.

% The dynSmag model is significantly cheaper, both in terms of FLOPs and evaluation methodology. For classical eddy viscosity models like dynSmag, the evaluation cost is typically measured per node rather than per quadrature point.

\vspace{3pt}\noindent{\underline{Key Takeaways}}

The SR model achieves significant reductions in both training and inference cost compared to the TBNN, owing to its reduced dimensionality, simpler optimization procedure, and streamlined parametric form. These efficiency gains are reflected in a training routine that is roughly an order of magnitude faster and uses less than a third of the peak memory, and in inference evaluations that require fewer than half as many FLOPs. Despite these reductions in computational cost, the SR models achieved predictive performance comparable to the TBNN across most of the metrics examined in Sections~\ref{sec:alpha}, \ref{sec:beta}, and \ref{sec:periodic hill}. Furthermore, note that the TBNN architecture and training dataset considered here are extremely small relative to typical implementations in the literature, where deep networks with substantially more layers and neurons are trained on vast libraries of data. Such increases in network capacity and dataset volume would require dramatically more computational resources, making the relative efficiency advantage of the SR framework proposed here even more pronounced.

%% ---------------------------
%% Conclusions
%% ---------------------------
\section{Conclusions and Future Work}\label{sec:conclusions}

In this work, we present a sparse regression-based SGS stress closure modeling framework with embedded invariance properties that accounts for filter anisotropy and incorporates dissipation-aware regularization. We leverage sparse regression to yield explicit algebraic representations of the SGS stress tensor while retaining only the most relevant terms, ensuring a compact and physically meaningful model. This enables the discovery of explicit polynomial models that require fewer computational resources to train and lower inference cost compared to more commonly used neural network-based turbulence closures. Physical symmetry and invariance properties are embedded directly into the model form by constructing the SGS stresses on a complete and minimal nondimensionalized tensor integrity basis derived from Smith’s representation theory \cite{Smith71} with exclusive dependence on spatial velocity gradients and grid- and fluid-based quantities. This formulation enforces Galilean, unit, reflectional, and rotational invariance, thereby ensuring adherence to the symmetries of the governing equations and enhancing generalization capabilities across distinct flow configurations \cite{Ling16}. Filter anisotropy awareness is incorporated through the utilization of a mapping between the anisotropic physical space and an isotropic parent space by following the approach of \cite{Prakash24}. This allows the discovered closures to remain accurate across anisotropic filtering operations and grid resolutions, further supplementing the generalizability of the models. Furthermore, we introduce an additional term to the elastic net objective function that penalizes SGS dissipation errors during the fitting process. This penalty term targets a common source of numerical instability in structural SGS models \citep{Vreman97}, encouraging the learned closure to retain more accurate dissipative behavior in deployment and thereby improve functional performance. Through these combined features, the resulting SGS models maintain tractability, generalizability, and stability while requiring only modest amounts of high-fidelity training data.

The results demonstrate that sparse polynomial closures identified through the proposed framework are capable of reproducing key SGS behaviors while maintaining a simple and computationally efficient representation. \textit{A priori} analysis indicates that increasing the predictor set order beyond $K=3$ provides no measurable benefit, suggesting that the dominant SGS dynamics can be captured with relatively low-order functional forms. While the TBNN outperformed the SR model in this baseline evaluation, subsequent \textit{a posteriori} tests --- for forced HIT at $Re_\lambda = \infty$, Taylor-Green vortex flow at $Re = 1,\!600$, and turbulent channel flow at $Re_\tau = 590$ --- demonstrate that tuning alternative parameters in the training objective can reduce the performance gap across different flow regimes. 

Varying the regularization parameter $\alpha$ alters the coefficient magnitudes and sparsity of the resulting closures. Analyses reveal that decreased model regularization does not consistently improve predictive accuracy. In some regimes, minimally regularized models produce the best agreement with reference data, whereas models with more regularization perform best in others. Adjusting the dissipation parameter $\beta$ penalizes deviations between the modeled and exact dissipation. Modifications to this penalty act to redistribute dissipative behavior across scales, altering the energetic behavior in the cascade. Larger penalties often disrupt intermediate-scale energy transfer, reducing high-wavenumber energy accumulation but frequently at the cost of over-dissipation. Smaller values better preserve inertial-range dynamics, but often allow excess small-scale energy pile-up. Across these variations in both $\alpha$ and $\beta$, several sparse regression models achieve predictive accuracy comparable to, and in some cases exceeding, that of the TBNN despite their simpler functional form. These results highlight the ability of the proposed framework to identify sparse closures that generalize across distinct flow conditions, geometric configurations, and filter resolutions beyond those encountered during training. This predictive capability extends to the more challenging separated flow benchmark of the periodic hill at $Re_h = 5,600$, where a representative sparse regression model achieves accuracy broadly comparable to the TBNN, and consistently outperforms the dynamic Smagorinsky model.

In addition to predictive performance, the sparse regression framework offers computational advantages. Relative to the TBNN, the training routine of the representative SR model is roughly an order of magnitude faster and requires less than a third of the peak memory, while inference evaluations involve fewer than half the floating-point operations. These results highlight the practical advantage of sparse polynomial data-driven model forms in scenarios where both predictive accuracy and computational tractability are critical.

Since the objective of this study is to establish a general SGS closure modeling framework rather than to finalize a single optimal model, the closures examined here represent an early realization with ample room for further refinement and enhancements. Namely, the \textit{a posteriori} results indicate that training exclusively on an idealized turbulence dataset limits the range of turbulent behaviors captured by the model and may not be representative of the diverse range of energy transfer characteristics present in other flow regimes, such as wall-bounded flows or those with non-equilibrium dynamics. Incorporating a broader set of training cases may yield even better predictive performance across these flow regimes. We also aim to extend this framework to convective flows, which involves modeling both the SGS stress tensor and heat flux vector to capture the coupled momentum and thermal transport. Furthermore, adopting a sparse regression-based strain-rate eigenframe \citep{Prakash22} approach would allow each component of the SGS tensor (or vector) in the strain-rate eigenframe to be modeled and tuned separately, providing additional flexibility in representing complex anisotropic effects and enabling more targeted enhancements. Finally, adapting the framework to model Reynolds stresses in RANS would expand its applicability to a broader range of turbulence problems across various fields of study, enabling the deployment of sparse efficient closures in more computationally accessible and tractable settings.

%% ---------------------------
%% Acknowledgements
%% ---------------------------
\section*{Acknowledgements} \label{sec:acknowledge}

This research was supported by the U.S. Department of Energy (DOE) Office of Science's Advanced Scientific Computing Research (ASCR) program under Award No. DE-SC0025656. The first author also acknowledges support from the Department of Defense (DOD) National Defense Science and Engineering Graduate Fellowship (NDSEG) program. The second author is supported by the Laboratory Directed Research and Development (LDRD) program and the Advanced Simulation and Computing (ASC) program at Los Alamos National Laboratory. The authors collectively acknowledge Prof. Kenneth E. Jansen for providing access to software and high-performance computing resources used in this work.

An award of computer time was provided by the U.S. DOE’s Innovative and Novel Computational Impact on Theory and Experiment (INCITE) Program. This research used resources from the Argonne Leadership Computing Facility, a U.S. DOE Office of Science user facility at Argonne National Laboratory, which is supported by the Office of Science of the U.S. DOE under Contract No. DE-AC02-06CH11357. This work also utilized the Alpine high performance computing resource at the University of Colorado Boulder. Alpine is jointly funded by the University of Colorado Boulder, the University of Colorado Anschutz, Colorado State University, and the National Science Foundation under Award No. 2201538.

%% ---------------------------
%% Bibliography and postamble
%% ---------------------------
%%%%%%%%%%%%%%%%%%%%%%%%%%%%%%%%%%%%%%%%%%%%%%%%%%%%%%%%%%%%%%%%%%%%%%%%%%
%% This file was autogenerated by PaperShell v2.6.1 on 2023-02-03 13:51:10
%% https://github.com/sylvainhalle/PaperShell
%% DO NOT EDIT!
%%%%%%%%%%%%%%%%%%%%%%%%%%%%%%%%%%%%%%%%%%%%%%%%%%%%%%%%%%%%%%%%%%%%%%%%%%
\bibliographystyle{elsarticle-num}

% \section*{References}

\bibliography{paper}
%%%%%%%%%%%%%%%%%%%%%%%%%%%%%%%%%%%%%%%%%%%%%%%%%%%%%%%%%%%%%%%%%%%%%%%%%%
%% This file was autogenerated by PaperShell v2.6.1 on 2023-02-03 13:51:10
%% https://github.com/sylvainhalle/PaperShell
%% DO NOT EDIT!
%%%%%%%%%%%%%%%%%%%%%%%%%%%%%%%%%%%%%%%%%%%%%%%%%%%%%%%%%%%%%%%%%%%%%%%%%%
%% ---------------------------
%% If there is anything you would like to insert *after* the bibliography
%% (and *before* the \end{document} instruction), place it here
%% ---------------------------

\appendix

\section{Definition of the Local Anisotropy Tensor}\label{sec:appendix -- aniso local}

In the implicitly filtered LES framework adopted here, the filter geometry is defined in relation to the underlying computational-grid geometry, such that the filter widths $\varDelta_i$ and principal filtering directions $\boldsymbol{a}_i$ are defined from the grid dimensions and orientations. The formulation presented in Section~\ref{sec:aniso} considers a globally defined filtering geometry and corresponding spatially constant anisotropy tensor $\boldsymbol{A}$. This setting is naturally realized for spatially uniform grid geometries, as for the structured Cartesian grid depicted in Figure~\ref{fig:anisotropic grid mapping}. For computational grids with spatially varying dimensions and orientations, however, such a global definition for $\boldsymbol{A}$ may not exist. Here, we provide a procedure for defining the local filtering geometry from the underlying grid geometry, from which the local anisotropy tensor $\boldsymbol{A}(\boldsymbol{x})$ is obtained.

\begin{figure}[b]
\centerline{\includegraphics[width=0.99\textwidth]{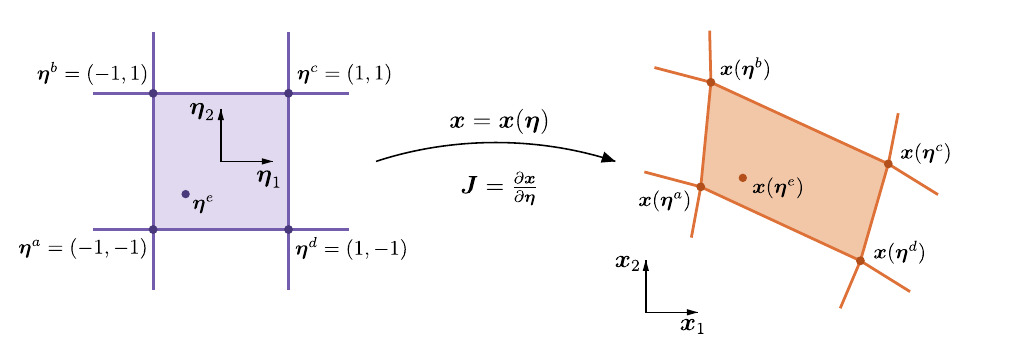}}% Images in 100% size
  \caption{Mapping from the square bi-unit reference cell in computational coordinates $\boldsymbol{\eta}$ to a general quadrilateral cell in physical coordinates $\boldsymbol{x}$.}
    \label{fig:computational grid mapping}
\end{figure}

To characterize the local grid geometry, consider a cell-wise invertible mapping between computational coordinates $\boldsymbol{\eta}$ on the bi-unit reference element, $[-1, 1]^d$, and physical coordinates $\boldsymbol{x}$, given by $\boldsymbol{x}=\boldsymbol{x}(\boldsymbol{\eta})$. This mapping is illustrated in Figure~\ref{fig:computational grid mapping}. Here, $\boldsymbol{x}(\boldsymbol{\eta})$ denotes the standard bilinear (trilinear) mapping for quadrilateral (hexahedral) cells; an analogous construction may be applied to other cell types. The Jacobian $\boldsymbol{J} = \partial \boldsymbol{x}/\partial \boldsymbol{\eta}$ provides the local linearization of the mapping between the computational and physical spaces. On the bi-unit reference element, we define the metric tensor as

\begin{equation}
    g_{ij} = 
        \frac{1}{4} 
        \frac{\partial \eta_k}{\partial x_i} 
        \frac{\partial \eta_k}{\partial x_j},
    \label{eqn:metric tensor}
\end{equation}

\noindent or, equivalently, $\boldsymbol{g} = \frac{1}{4} \boldsymbol{J}^{-T} \boldsymbol{J}^{-1}$. The metric tensor provides a measure of the local geometric stretching associated with the mapping between the computational and physical spaces. Because $\boldsymbol{g}$ is symmetric by construction, it admits an eigendecomposition of the form

\begin{equation}
    \boldsymbol{g} = 
        \boldsymbol{Q} 
        \boldsymbol{\Lambda} 
        \boldsymbol{Q}^{T},
    \label{eqn:eigendecomp}
\end{equation}

\noindent where the eigenvector columns $\boldsymbol{q}_i$ of $\boldsymbol{Q}$ give the local principal directions of the grid geometry, while the corresponding eigenvalues $\lambda_i$ in $\boldsymbol{\Lambda}$ characterize the geometric stretching along these directions. We assume that these local geometric quantities define a local ellipsoidal filter, with local principal filtering directions $\boldsymbol{a}_i(\boldsymbol{x}) = \boldsymbol{q}_i(\boldsymbol{x})$ and local filter widths $\varDelta_i(\boldsymbol{x}) = \lambda_i(\boldsymbol{x})^{-1/2}$. The resulting local filter width tensor is 

\begin{equation}
    \boldsymbol{\varDelta}(\boldsymbol{x}) =
        \sum_{i=1}^{d}
            \varDelta_i(\boldsymbol{x})\, \boldsymbol{a}_i(\boldsymbol{x})
            \otimes \boldsymbol{a}_i(\boldsymbol{x}),
    \label{eqn:Delta local}
\end{equation}

\noindent from which the anisotropy tensor can be defined locally as

\begin{equation}
    \boldsymbol{A}(\boldsymbol{x}) =
        \sqrt{d}\,
        \frac{\boldsymbol{\varDelta}(\boldsymbol{x})}
        {\left\|\boldsymbol{\varDelta}(\boldsymbol{x})\right\|_F}.
    \label{eqn:A local}
\end{equation}

\noindent The locally defined filter widths and principal directions are taken to characterize an ellipsoidal filter at each point in the domain, with its anisotropy described by the local anisotropy tensor $\boldsymbol{A}(\boldsymbol{x})$.

The closure modeling framework presented in this work is developed under the spatially constant anisotropy formulation of Section~\ref{sec:aniso}, for which the anisotropy tensor $\boldsymbol{A}$ is defined globally. The closure models are also learned in this setting. Upon inference, however, the models are applied locally using the spatially varying anisotropy tensor $\boldsymbol{A}(\boldsymbol{x})$ defined here. As noted previously, a spatially varying filter width can give rise to commutation errors, which are not accounted for here. The treatment of these effects in the local formulation is left for future development of the framework. For a spatially uniform grid geometry, $\boldsymbol{A}(\boldsymbol{x})$ reduces everywhere to the constant global anisotropy tensor $\boldsymbol{A}$, recovering the setting considered in Section~\ref{sec:aniso}.

% \newpage

\section{Construction of Truncated Predictor Set}\label{sec:appendix -- predictor}

To construct the predictor sets, we begin by notating the degree of each invariant scalar within Smith's basis:

\begin{equation}
   \Lambda = \{ \underarrow[1_{}][\uparrow]{\begin{math}k=0\end{math}} , \qquad
                              \underarrow[\lambda_0][\uparrow]{\begin{math}k=2\end{math}} , \qquad
                              \underarrow[\lambda_1][\uparrow]{\begin{math}k=2\end{math}} , \qquad
                              \underarrow[\lambda_2][\uparrow]{\begin{math}k=3\end{math}} , \qquad
                              \underarrow[\lambda_3][\uparrow]{\begin{math}k=3\end{math}} , \qquad
                              \underarrow[\lambda_4][\uparrow]{\begin{math}k=4\end{math}} , \qquad
                              \underarrow[\lambda_5][\uparrow]{\begin{math}k=1\end{math}} \},
    \label{eqn:lambda degrees}
\end{equation}

\noindent where $k$ denotes the degree and $\{\boldsymbol{\cdot}\}$ denotes a set. We added the scalar $\mathrm{1}$ to the set of invariant scalars, $\Lambda$, to account for cases in which a tensor basis's scaling coefficient is independent of $\hat{\tilde{S}}_{ij}$, $\hat{\tilde{\varOmega}}_{ij}$, and $\hat{\nu}$. This reasoning will become more clear in a few steps. Using the above degree specifications, we can define sets of the scalars based on their degree

\noindent
\begin{minipage}{0.33\textwidth}
    \begin{align}
        &\Lambda^{\{ k=0 \}} = \{ 1 \} \nonumber \\
        &\Lambda^{\{ k=1 \}} = \{ \lambda_5 \} \nonumber
    \end{align}
\end{minipage}
\hfill
\begin{minipage}{0.32\textwidth}
    \begin{align}
        &\Lambda^{\{ k=2 \}} = \{ \lambda_0 , \lambda_1 \} \nonumber \\
        &\Lambda^{\{ k=3 \}} = \{ \lambda_2 , \lambda_3 \} \nonumber
    \end{align}
\end{minipage}
\hfill
\begin{minipage}{0.33\textwidth}
    \begin{align}
    \begin{split}
        &\Lambda^{\{ k=4 \}} = \{ \lambda_4 \}. \\
        & 
    \end{split}
    \label{eqn:lambda degree sets}
    \end{align}
\end{minipage} \bigskip

\noindent Following the same procedure for the set of basis tensors, $\Tau$, gives

\begin{equation}
    \Tau = \{ \underarrow[\boldsymbol{T}^{(0)}][\uparrow]{\begin{math}k=0\end{math}} , \quad
                          \underarrow[\boldsymbol{T}^{(1)}][\uparrow]{\begin{math}k=1\end{math}} , \quad
                          \underarrow[\boldsymbol{T}^{(2)}][\uparrow]{\begin{math}k=2\end{math}} , \quad
                          \underarrow[\boldsymbol{T}^{(3)}][\uparrow]{\begin{math}k=2\end{math}} , \quad
                          \underarrow[\boldsymbol{T}^{(4)}][\uparrow]{\begin{math}k=2\end{math}} , \quad
                          \underarrow[\boldsymbol{T}^{(5)}][\uparrow]{\begin{math}k=3\end{math}} , \quad
                          \underarrow[\boldsymbol{T}^{(6)}][\uparrow]{\begin{math}k=3\end{math}} , \quad
                          \underarrow[\boldsymbol{T}^{(7)}][\uparrow]{\begin{math}k=4\end{math}} \}
    \label{eqn:T degrees}
\end{equation}

\noindent and

\noindent
\begin{minipage}{0.33\textwidth}
    \begin{align}
        &\Tau^{\{ k=0 \}} = \{ \boldsymbol{T}^{(0)} \} \nonumber \\
        &\Tau^{\{ k=1 \}} = \{ \boldsymbol{T}^{(1)} \} \nonumber
    \end{align}
\end{minipage}
\hfill
\begin{minipage}{0.32\textwidth}
    \begin{align}
        &\Tau^{\{ k=2 \}} = 
            \{ \boldsymbol{T}^{(2)} , 
               \boldsymbol{T}^{(3)} , 
               \boldsymbol{T}^{(4)} \} \nonumber \\
        &\Tau^{\{ k=3 \}} =
            \{ \boldsymbol{T}^{(5)} , 
               \boldsymbol{T}^{(6)} \} \nonumber
    \end{align}
\end{minipage}
\hfill
\begin{minipage}{0.33\textwidth}
    \begin{align}
    \begin{split}
        &\Tau^{\{ k=4 \}} = \{ \boldsymbol{T}^{(7)} \}. \\
        & 
    \end{split}
    \label{eqn:T degree sets}
    \end{align}
\end{minipage} \bigskip

Using Equations~\eqref{eqn:lambda degree sets} and \eqref{eqn:T degree sets}, we can define predictor sets of degree $k$, $P^{\{ k\}}$, through an appropriate scaling of particular basis tensor sets by sets of invariant scalars. The predictor set of degree zero, for example, is defined as the multiplication of the set of degree-zero scalars with the set of degree-zero tensors

\begin{flalign} 
    \begin{aligned}
        P^{\{ k=0 \}} 
            &= \Lambda^{\{ k=0 \}} \Tau^{\{ k=0 \}} \\ 
            &= \{ 1  \boldsymbol{T}^{(0)} \}. \\
    \end{aligned} &&
    \label{eqn:P0 degree sets}
\end{flalign} 

\noindent Sets of $P$ with higher degrees necessitate finding the union of two or more sets of scaled basis tensors. $P^{\{ k=1 \}}$, for example, is constructed by finding the union ($\cup$) of degree-zero tensors scaled by degree-one scalars ($\Lambda^{\{ k=1 \}} \Tau^{\{ k=0 \}}$) and degree-one tensors scaled by degree-zero scalars ($\Lambda^{\{ k=0 \}} \Tau^{\{ k=1 \}}$) as follows:

\begin{flalign} 
    \begin{aligned}
        P^{\{ k=1 \}}
            &= \Lambda^{\{ k=1 \}} \Tau^{\{ k=0 \}} \cup
               \Lambda^{\{ k=0 \}} \Tau^{\{ k=1 \}} \\ 
            &= \{ \lambda_5 \boldsymbol{T}^{(0)}, 
                  1  \boldsymbol{T}^{(1)} \}. \\ 
    \end{aligned} &&
    \label{eqn:P1 degree sets}
\end{flalign}

\noindent Higher degree sets have increasing quantities of scaled basis tensors within each set

\begin{flalign} 
    \begin{aligned}
        P^{\{ k=2 \}}
            &= \Lambda^{\{ k=2 \}} \Tau^{\{ k=0 \}} \cup 
               \Lambda^{\{ k=1 \}^2} \Tau^{\{ k=0 \}} \cup 
               \Lambda^{\{ k=1 \}} \Tau^{\{ k=1 \}} \cup 
               \Lambda^{\{ k=0 \}} \Tau^{\{ k=2 \}} \\
            &= \{ \lambda_0  \boldsymbol{T}^{(0)} , \lambda_1  \boldsymbol{T}^{(0)} , 
                  \lambda_5^2  \boldsymbol{T}^{(0)} , 
                  \lambda_5 \boldsymbol{T}^{(1)} , 
                  1\boldsymbol{T}^{(2)} , 1\boldsymbol{T}^{(3)} , 1\boldsymbol{T}^{(4)} \}, \\
    \end{aligned} &&
    \label{eqn:P2 degree sets}
\end{flalign}

\begin{flalign} 
    \begin{aligned}
        P^{\{ k=3 \}}
            &= \Lambda^{\{ k=3 \}} \Tau^{\{ k=0 \}} \cup
               \Lambda^{\{ k=1 \}} \Lambda^{\{ k=2 \}} \Tau^{\{ k=0 \}} \cup
               \Lambda^{\{ k=1 \}^3} \Tau^{\{ k=0 \}} \cup
               \Lambda^{\{ k=2 \}} \Tau^{\{ k=1 \}} \\ &\qquad \cup 
               \Lambda^{\{ k=1 \}^2} \Tau^{\{ k=1 \}} \cup
               \Lambda^{\{ k=1 \}} \Tau^{\{ k=2 \}} \cup
               \Lambda^{\{ k=0 \}} \Tau^{\{ k=3 \}} \\
            &= \{ \lambda_2  \boldsymbol{T}^{(0)} , \lambda_3  \boldsymbol{T}^{(0)} , 
                  \lambda_5 \lambda_0 \boldsymbol{T}^{(0)} , \lambda_5 \lambda_1 \boldsymbol{T}^{(0)} , 
                  \lambda_5^3 \boldsymbol{T}^{(0)} ,
                  \lambda_0  \boldsymbol{T}^{(1)} , \lambda_1  \boldsymbol{T}^{(1)} , \\ &\qquad 
                  \lambda_5^2  \boldsymbol{T}^{(1)} , 
                  \lambda_5 \boldsymbol{T}^{(2)} , \lambda_5 \boldsymbol{T}^{(3)} , 
                        \lambda_5 \boldsymbol{T}^{(4)} ,
                  1\boldsymbol{T}^{(5)} , 1\boldsymbol{T}^{(6)} \}, \\
    \end{aligned} &&
    \label{eqn:P3 degree sets}
\end{flalign}

\begin{flalign} 
    \begin{aligned}
        P^{\{ k=4 \}}
            &= \Lambda^{\{ k=4 \}} \Tau^{\{ k=0 \}} \cup
               \Lambda^{\{ k=3 \}} \Lambda^{\{ k=1 \}} \Tau^{\{ k=0 \}} \cup
               \Lambda^{\{ k=2 \}^2} \Tau^{\{ k=0 \}} \cup
               \Lambda^{\{ k=2 \}} \Lambda^{\{ k=1 \}^2} \Tau^{\{ k=0 \}} \\ &\qquad \cup
               \Lambda^{\{ k=1 \}^4} \Tau^{\{ k=0 \}} \cup 
               \Lambda^{\{ k=3 \}} \Tau^{\{ k=1 \}} \cup 
               \Lambda^{\{ k=2 \}} \Lambda^{\{ k=1 \}} \Tau^{\{ k=1 \}} \cup
               \Lambda^{\{ k=1 \}^3} \Tau^{\{ k=1 \}} \\ &\qquad \cup
               \Lambda^{\{ k=2 \}} \Tau^{\{ k=2 \}} \cup
               \Lambda^{\{ k=1 \}^2} \Tau^{\{ k=2 \}} \cup
               \Lambda^{\{ k=1 \}} \Tau^{\{ k=3 \}} \cup
               \Lambda^{\{ k=0 \}} \Tau^{\{ k=4 \}} \\
            &= \{ \cdots \}. \\
    \end{aligned} &&
    \label{eqn:P4 degree sets}
\end{flalign}

\noindent We have only provided here the predictor sets up to a degree of four, but note that this same procedure can be followed for arbitrarily high input degrees.

\section{Sparse Regression Models --- Predictors and Coefficients} \label{sec:appendix -- model expressions}

Table~\ref{tab:SR models} reports the predictors $\boldsymbol{P}_n$ and their corresponding coefficients $w_n$ downselected by the sparse regression algorithm for the models analyzed in this work. Columns correspond to individual model configurations indexed by model order $K$, regularization penalty $\alpha$, and the dissipation penalty $\beta$, while rows enumerate the retained predictors grouped polynomial order.

% \begin{landscape}
\begin{sidewaystable}[p]
\centering

\renewcommand{\arraystretch}{1.18}
% \begin{table}[!ht]
    % \centering
    % \resizebox{\linewidth}{!}{
    \NiceMatrixOptions{custom-line={command = DoubleRule , multiplicity = 2 , sep-color = white}}
    % \NiceMatrixOptions{hvlines-except-borders}
    \resizebox{\textheight}{!}{%
    \begin{NiceTabular}{c|c|d|d|d|d|d|d|d|d|d}
    % \begin{NiceTabular}{c|c|c|c|c|c|c|c|c|c|c}
    \DoubleRule
            \Block{4-1}{$k$}
                & \Block{4-1}{$\boldsymbol{P}_n$} 
                & \Block{1-9}{$w_n/10^{2}$} \\ \cline{3-11}
            &   & \multicolumn{1}{c}{$K=2$}         & \multicolumn{1}{c}{$K=3$}         
                & \multicolumn{1}{c}{$K=3$}         & \multicolumn{1}{c}{$K=3$}         
                & \multicolumn{1}{c}{$K=3$}         & \multicolumn{1}{c}{$K=3$}         
                & \multicolumn{1}{c}{$K=3$}         & \multicolumn{1}{c}{$K=4$}         
                & \multicolumn{1}{c}{$K=5$} \\ \cline{3-11}
            &   & \multicolumn{1}{c}{$\alpha=1e-4$} & \multicolumn{1}{c}{$\alpha=1e-3$} 
                & \multicolumn{1}{c}{$\alpha=1e-4$} & \multicolumn{1}{c}{$\alpha=1e-4$} 
                & \multicolumn{1}{c}{$\alpha=1e-4$} & \multicolumn{1}{c}{$\alpha=1e-4$} 
                & \multicolumn{1}{c}{$\alpha=1e-5$} & \multicolumn{1}{c}{$\alpha=1e-4$} 
                & \multicolumn{1}{c}{$\alpha=1e-4$} \\ \cline{3-11}
            &   & \multicolumn{1}{c}{$\beta=0$}     & \multicolumn{1}{c}{$\beta=0$}     
                & \multicolumn{1}{c}{$\beta=0$}     & \multicolumn{1}{c}{$\beta=0.5$}   
                & \multicolumn{1}{c}{$\beta=1$}     & \multicolumn{1}{c}{$\beta=1.5$}   
                & \multicolumn{1}{c}{$\beta=0$}     & \multicolumn{1}{c}{$\beta=0$}     
                & \multicolumn{1}{c}{$\beta=0$}  \\ \hline
            \Block{1-1}{$0$} 
            & $\boldsymbol{T}^{(0)}$
                & 4.20211       & 6.02649	    & 4.18957	    & 4.27671	    & 4.34398	  
                                & 4.42463       & 3.15162	    & 4.34896	    & 4.35241 \\ \hline
            \Block{2-1}{$1$} 
            & $\lambda_5 \boldsymbol{T}^{(0)}$
                & -26.47555	    & -10.99720	    & -30.29164	    & -31.25109	    & -32.16950	
                                & -33.10890	    & -36.17821	    & -30.47281	    & -30.56706 \\ \cline{2-11}
            & $\boldsymbol{T}^{(1)}$
                & -1.64655	    & -1.20504	    & -1.78912	    & -1.92450	    & -2.04561	
                                & -2.17848	    & -2.20748	    & -1.78199	    & -1.77709 \\ \hline
            \Block{7-1}{$2$} 
            & $\lambda_0 \boldsymbol{T}^{(0)}$
                & 2.13718	    &         	    & 2.44866	    & 2.35963	    & 2.30459	
                                & 2.22319	    & 3.72345	    & 2.43772	    & 2.43044 \\ \cline{2-11}
	        & $\lambda_1 \boldsymbol{T}^{(0)}$
                &	            &	            &	            &	            &	
                                &	            & -0.57528	    &	            &            \\ \cline{2-11}
	        & $\lambda_5^2 \boldsymbol{T}^{(0)}$
                & 15.72324	    &               & 26.20500	    & 28.26680	    & 30.34600	
                                & 32.48222	    & 55.99747	    & 24.33578	    & 24.52194 \\ \cline{2-11}
	        & $\lambda_5 \boldsymbol{T}^{(1)}$
                & 3.07160	    &               & 3.22545	    & 3.59713	    & 3.98177	
                                & 4.35521	    & 8.09298	    & 3.23952	    & 3.21125 \\ \cline{2-11}
	        & $\boldsymbol{T}^{(2)}$
                & 6.08836	    & 3.72640	    & 6.14268	    & 7.64034	    & 9.15336	
                                & 10.65928	    & 7.05363	    & 5.99680	    & 6.03284 \\ \cline{2-11}
	        & $\boldsymbol{T}^{(3)}$
                & -7.12394	    &	            & -7.10431	    & -6.99165	    & -6.89849	
                                & -6.80117	    & -8.85617	    & -7.12837	    & -7.13276 \\ \cline{2-11}
	        & $\boldsymbol{T}^{(4)}$
                & -8.09557	    & -6.34003	    & -8.90761	    & -9.07001	    & -9.23547	
                                & -9.39056	    & -9.65466	    & -8.90237	    & -8.90582 \\ \hline
            \Block{10-1}{$3$} 
            & $\lambda_2 \boldsymbol{T}^{(0)}$
                &	            &	            & 1.31572	    & 1.29594	    & 1.26853	
                                & 1.27931	    & 1.67273	    & 1.30100	    & 1.30782 \\ \cline{2-11}
	        & $\lambda_3 \boldsymbol{T}^{(0)}$
                &	            &	            &	            &	            &	
                                &	            & -1.40684	    &	            &            \\ \cline{2-11}
	        & $\lambda_0 \lambda_5 \boldsymbol{T}^{(0)}$
                &	            &	            &	            & -0.15995	    & -0.48468	
                                & -0.72538	    & -5.79397	    & -0.35001	    & -0.33441 \\ \cline{2-11}
	        & $\lambda_5^3 \boldsymbol{T}^{(0)}$
                &	            &	            & -5.10041	    & -5.68129	    & -6.22541	
                                & -6.86870	    & -17.34852	    &	            &            \\ \cline{2-11}
	        & $\lambda_1 \boldsymbol{T}^{(1)}$
                &	            &	            &	            &	            &	
                                &	            & 0.66738	    &	            &            \\ \cline{2-11}
	        & $\lambda_5^2 \boldsymbol{T}^{(1)}$
                &	            &	            &	            &	            &	
                                &	            & -4.84820	    &	            &            \\ \cline{2-11}
	        & $\lambda_5 \boldsymbol{T}^{(2)}$
                &	            &	            & -1.30467	    & -4.67033	    & -7.99300	
                                & -11.36717	    & -11.16727	    & -1.22302	    & -1.28631 \\ \cline{2-11}
	        & $\lambda_5 \boldsymbol{T}^{(3)}$
                &	            &	            &	            &	            &	
                                &	            & 16.35922	    &	            &            \\ \cline{2-11}
	        & $\lambda_5 \boldsymbol{T}^{(4)}$
                &	            &               & 11.45650	    & 11.73420	    & 12.05054	
                                & 12.30287	    & 19.21204	    & 11.48474	    & 11.45581  \\ \cline{2-11}
	        & $\boldsymbol{T}^{(5)}$
                &	            &	            &	            &	            &	
                                & 0.29888	    & 4.44245	    &	            &            \\ \hline
            \Block{3-1}{$4$}     
            & $\lambda_4 \boldsymbol{T}^{(0)}$
                &	            &	            &	            &	            &	
                                &	            &	            & 1.58776	    & 1.60851 \\ \cline{2-11}
	        & $\lambda_5^4 \boldsymbol{T}^{(0)}$
                &	            &	            &	            &	            &	
                                &	            &	            & -2.07520	    & -2.12644 \\ \cline{2-11}
	        & $\lambda_2 \boldsymbol{T}^{(1)}$
                &	            &	            &	            &	            &	
                                &	            &	            & 0.23343	    & 0.20941 \\ \hline
            \Block{1-2}{Total terms:}  &
                & \multicolumn{1}{c}{9}             & \multicolumn{1}{c}{5}
                & \multicolumn{1}{c}{13}            & \multicolumn{1}{c}{14}
                & \multicolumn{1}{c}{14}            & \multicolumn{1}{c}{15}
                & \multicolumn{1}{c}{20}            & \multicolumn{1}{c}{16}
                & \multicolumn{1}{c}{16} \\
        \hline\hline
    \end{NiceTabular}
    }
    \caption{SR Model predictors and coefficients}
    \label{tab:SR models}
% \end{table}
% \renewcommand{\arraystretch}{1}

% \pagebreak

% \end{landscape}
\end{sidewaystable}

% \noindent 

\section{Supplemental Results}\label{sec:appendix -- results}

% {\parfillskip=0pt
This section presents supplemental \textit{a posteriori} analyses that complement the results reported in the main text. Sections~\ref{sec:appendix -- alpha} and \ref{sec:appendix -- beta} examine the sensitivity of flow statistics to variations in the regularization penalty $\alpha$ and dissipation penalty $\beta$, respectively. For the Taylor-Green vortex flow, the analysis focuses on the three-dimensional energy spectra at a time near peak dissipation, while for the turbulent channel flow, the Reynolds stress tensor components are considered. Section~\ref{sec:appendix -- periodic hill} evaluates the normal and shear Reynolds stress component profiles at multiple streamwise stations along the periodic hill.
% \par}

\begin{figure}[b]
\centerline{\includegraphics[width=0.99\textwidth]{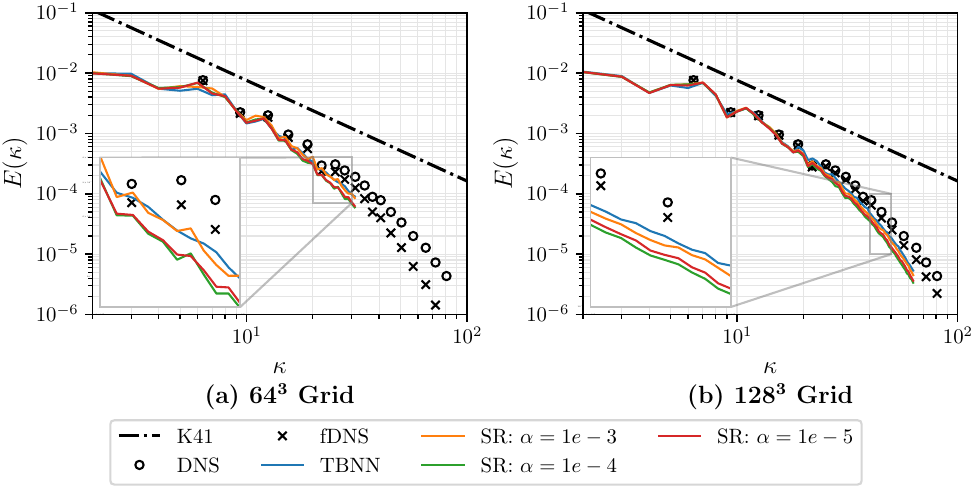}}% Images in 100% size
  \caption{Three-dimensional energy spectra at $t=9$ for Taylor-Green vortex flow at $Re = 1,600$ on (a) $64^3$ and (b) $128^3$ grid resolutions for models with varying Lasso penalty parameters ($\alpha$).}
    \label{fig:result - TGV spectra alpha}
\end{figure}

\subsection{\textit{A Posteriori} Sensitivity Analysis: Effect of Regularization Penalty ($\alpha$)} \label{sec:appendix -- alpha}

\begin{figure}[b]
\centerline{\includegraphics[width=0.99\textwidth]{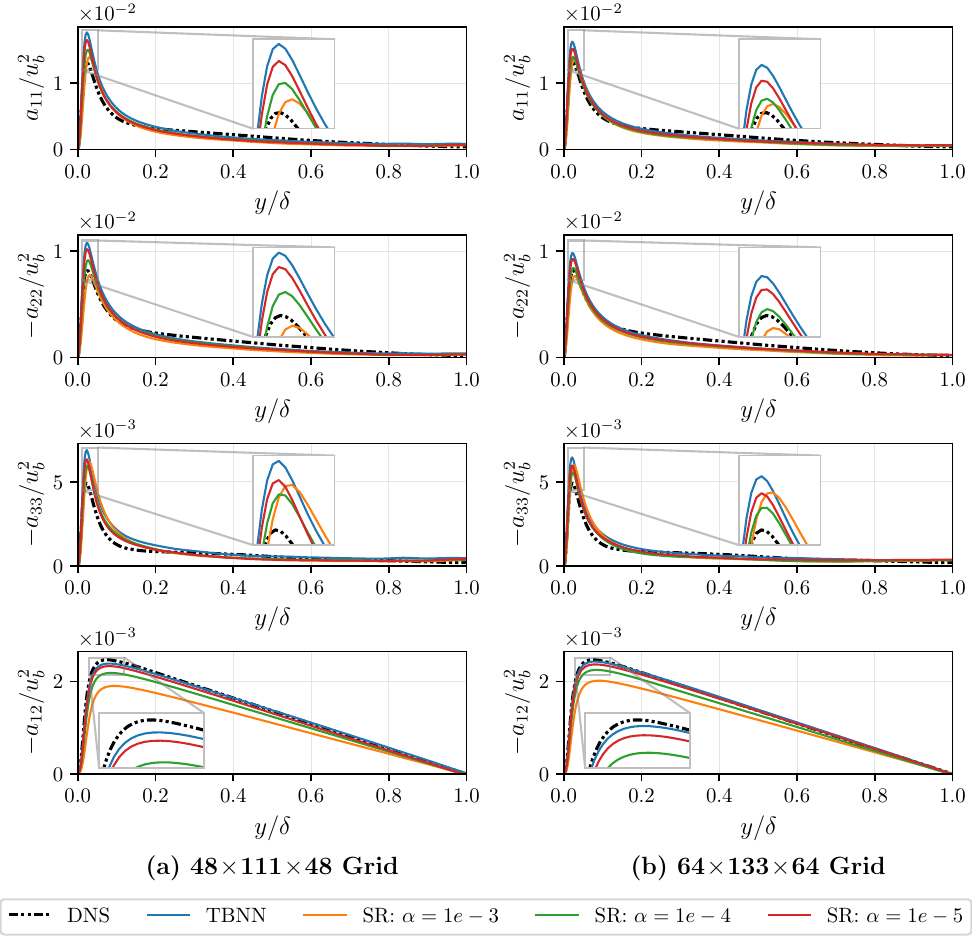}}% Images in 100% size
  \caption{Reynolds stress tensor components for turbulent channel flow at $Re_\tau = 590$ on the (a) $48\times111\times48$ and (b) $64\times133\times64$ grid resolutions for models with varying Lasso penalty parameters ($\alpha$).}
    \label{fig:result - Channel RS alpha}
\end{figure}

\vspace{3pt}\noindent{\underline{Taylor-Green Vortex Flow at $Re=1,600$}}

Figure~\ref{fig:result - TGV spectra alpha} shows the energy spectra recorded at t=9, a time near peak dissipation, for the Taylor-Green vortex flow for two different grid resolutions. At the larger scales, all the models exhibit strong agreement with each other and with the fDNS results, where agreement is sustained over a wider range of wavenumbers on the finer mesh. At the smaller scales, the SR model with $\alpha=1e-3$ achieves comparable performance to the TBNN, with prediction accuracy degrading as model regularization is reduced. This trend is consistent across both grid resolutions and with those observed in the resolved dissipation in Figure~\ref{fig:result - TGV eps alpha}.

\vspace{3pt}\noindent{\underline{Turbulent Channel Flow at $Re_\tau=590$}}

Select Reynolds stress component profiles for the turbulent channel flow are shown in Figure~\ref{fig:result - Channel RS alpha} for the (a) coarse and (b) medium grid resolutions. In the case of the streamwise direction ($a_{11}$), wall-normal direction ($a_{22}$), and spanwise-direction ($a_{33}$) normal Reynolds stress components, the TBNN significantly overpredicts the peak normal stress magnitudes across both grid resolutions. The peak normal stress magnitudes predicted by the SR models generally increase as $\alpha$ decreases, such that the more regularized models ($\alpha=1e-3$ and $\alpha=1e-4$) yield more accurate magnitude predictions. With stronger regularization, however, the location of the peak normal stress prediction shifts further away from the wall. In contrast with the normal components, the shear Reynolds stress component ($a_{12}$) predictions by the TBNN are closely aligned with the DNS and the SR model performance improves as regularization is reduced. Across all components and morels, predictions improve with grid refinement.

\subsection{\textit{A Posteriori} Sensitivity Analysis: Effect of Dissipation Penalty ($\beta$)} \label{sec:appendix -- beta}

\vspace{3pt}\noindent{\underline{Taylor-Green Vortex Flow at $Re=1,600$}}

\begin{figure}[b]
\centerline{\includegraphics[width=0.99\textwidth]{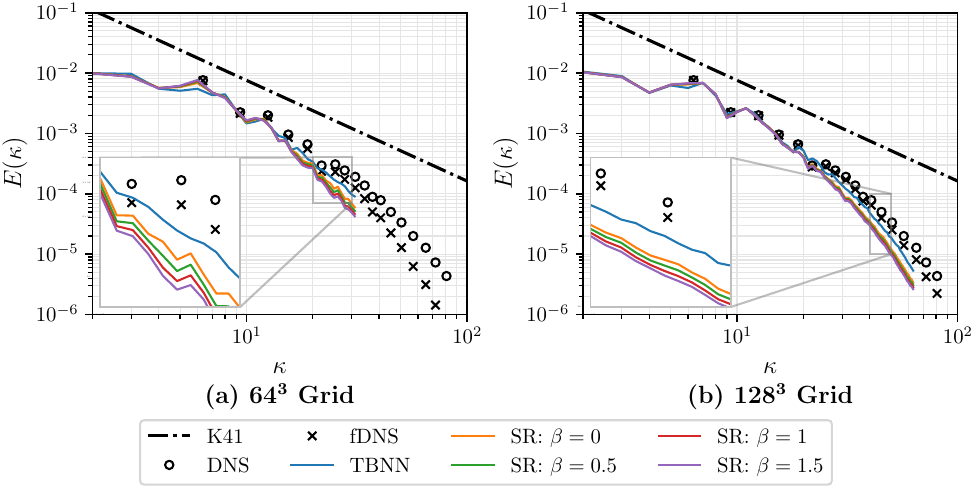}}% Images in 100% size
  \caption{Three-dimensional energy spectra at $t=9$ for Taylor-Green vortex flow at $Re = 1,600$ on (a) $64^3$ and (b) $128^3$ grid resolutions for models with varying penalty parameters on the SGS dissipation correction ($\beta$).}
    \label{fig:result - TGV spectra beta}
\end{figure}

In Figure~\ref{fig:result - TGV spectra beta}, we plot the three-dimensional energy spectra at $t=9$ for the Taylor-Green vortex flow. Model performance is consistent with the fDNS results at the larger scales. As the wavenumber increases, the TBNN deviates only mildly from the reference, whereas the SR models progressively underpredict the energy. Moreover, SR prediction errors grow with increasing $\beta$, reflecting that models with higher $\beta$ values are more dissipative on the smaller scales. This observation is consistent with the analysis of the resolved dissipation in Figure~\ref{fig:result - TGV eps beta}.

\begin{figure}[b!]
\centerline{\includegraphics[width=0.99\textwidth]{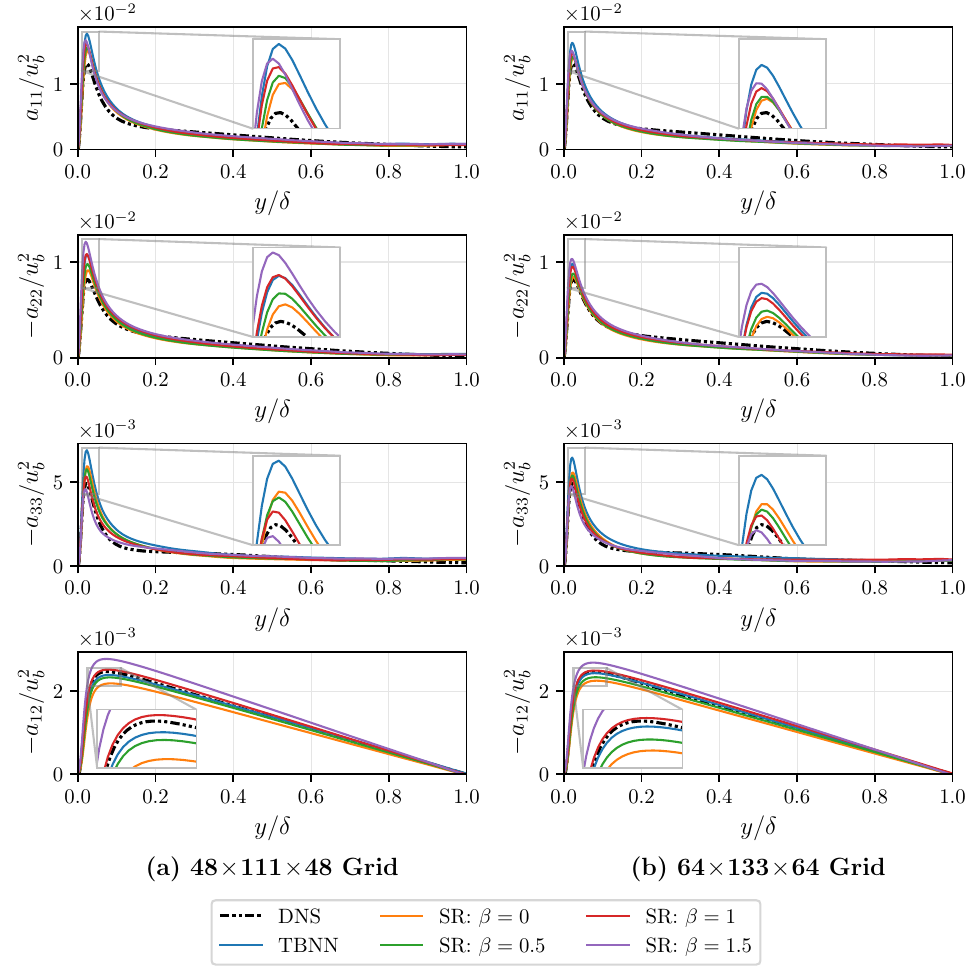}}% Images in 100% size
  \caption{Reynolds stress tensor components for turbulent channel flow at $Re_\tau = 590$ on the (a) $48\times111\times48$ and (b) $64\times133\times64$ grid resolutions for models with varying penalty parameters on the SGS dissipation correction ($\beta$).}
    \label{fig:result - Channel RS beta}
\end{figure}

\vspace{3pt}\noindent{\underline{Turbulent Channel Flow at $Re_\tau=590$}}

Figure~\ref{fig:result - Channel RS beta} shows select Reynolds stress component profiles for the Turbulent channel flow across two grid resolutions. For the $a_{11}$ and $a_{22}$ Reynolds stress components, as $\beta$ increases, the magnitude of the peak stress overprediction increases and the location of the peak stress prediction shifts closer to the wall. The same wallward-shift of the peak stress y-location with increasing $\beta$ is observed for the $a_{33}$ Reynolds stress component, but the magnitude of the peak stress overprediction is reduced as $\beta$ increases. In the $a_{12}$ component, increasing $\beta$ to intermediate values once again improves alignment with DNS peak stress predictions. The SR models generally outperform the TBNN across each Reynolds stress component and grid resolution.

\subsection{\textit{A Posteriori} Performance Benchmark: Separated Flow} \label{sec:appendix -- periodic hill}

\begin{figure}[b]
\centerline{\includegraphics[width=0.99\textwidth]{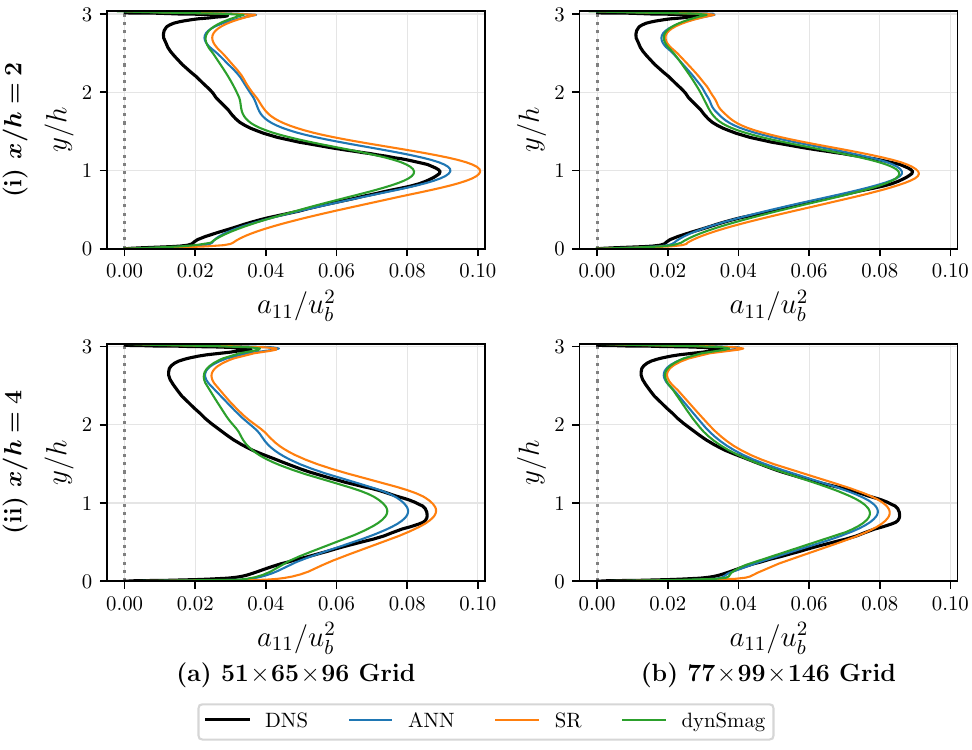}}% Images in 100% size
  \caption{Normal Reynolds stress component profiles in the streamwise direction for the periodic hill at $Re_h=5,600$ on (a) the coarser and (b) the finer grid resolutions at (\romannumeral1\relax) $x/h=2$ and (\romannumeral2\relax) $x/h=4$.}
    \label{fig:result - Hill RS uu}
\end{figure}

The normal Reynolds stress component profiles in the streamwise direction at (\romannumeral1\relax) $x/h=2$ and (\romannumeral2\relax) $x/h=4$ are shown in Figure~\ref{fig:result - Hill RS uu} for the (a) coarser and (b) finer grid resolutions. At both $x$-locations, the models predict excessively steep gradients of $a_{11}$ near the wall, leading to overshoots relative to DNS that are exacerbated on the coarser mesh. The SR model, in particular, underperforms in this regard. The predicted $y$-location of the peak streamwise-normal Reynolds stresses remains reasonable across models and resolutions, although peak magnitudes are captured with varying accuracy: at $x/h=2$, the TBNN provides the closest agreement on the coarser grid but is surpassed by the SR model on the finer grid, whereas at $x/h=4$, the SR model best represents the peak stresses for both resolutions. Furthermore, the models all overpredict the stress component magnitudes in the recovery region at both $x$-stations as the flow transitions back into the bulk.

\begin{figure}[b]
\centerline{\includegraphics[width=0.99\textwidth]{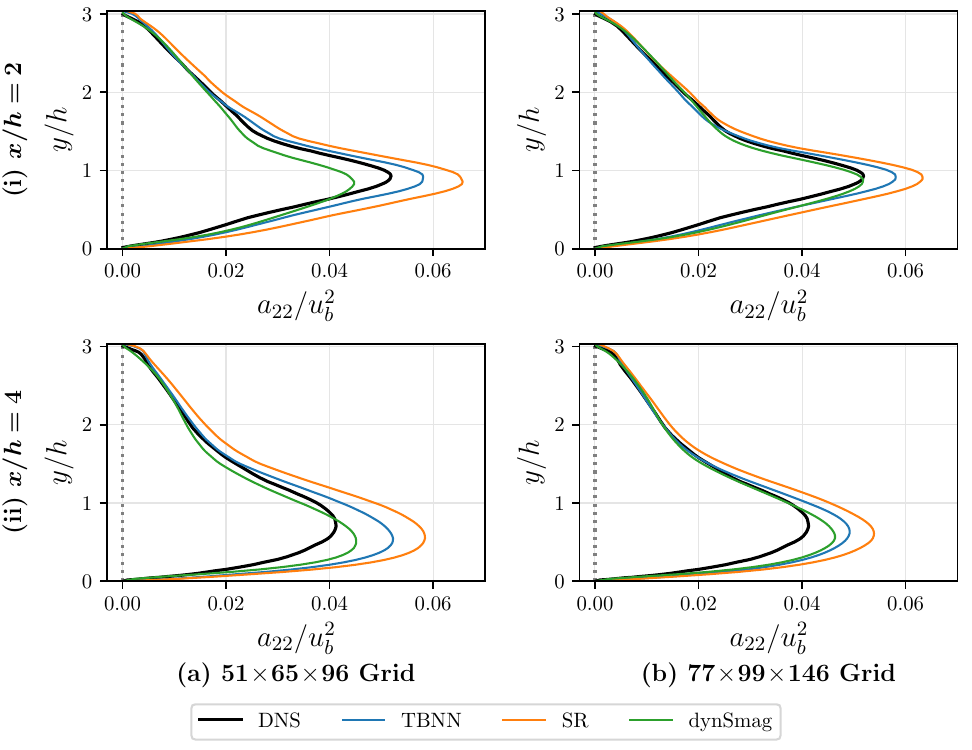}}% Images in 100% size
  \caption{Normal Reynolds stress component profiles in the wall-normal direction for the periodic hill at $Re_h=5,600$ on (a) the coarser and (b) the finer grid resolutions at (\romannumeral1\relax) $x/h=2$ and (\romannumeral2\relax) $x/h=4$.}
    \label{fig:result - Hill RS vv}
\end{figure}

Figure~\ref{fig:result - Hill RS vv} presents the normal Reynolds stress component in the wall-normal direction, $a_{22}$, at the same $x$-locations as Figure~\ref{fig:result - Hill RS uu}. At $x/h = 2$, the data-driven models overpredict the magnitude of the peak wall-normal-direction Reynolds stress component, but capture its $y$-position reasonably closely. In contrast, the dynSmag model gives close agreement to the DNS-predicted peak stress component magnitude on the finer mesh, though at a lower $y$-value. At $x/h=4$, all models overestimate the peak value of the $a_{22}$ profile and place the $y$-location of the peak closer to the wall than the reference.

\begin{figure}
\centerline{\includegraphics[width=0.99\textwidth]{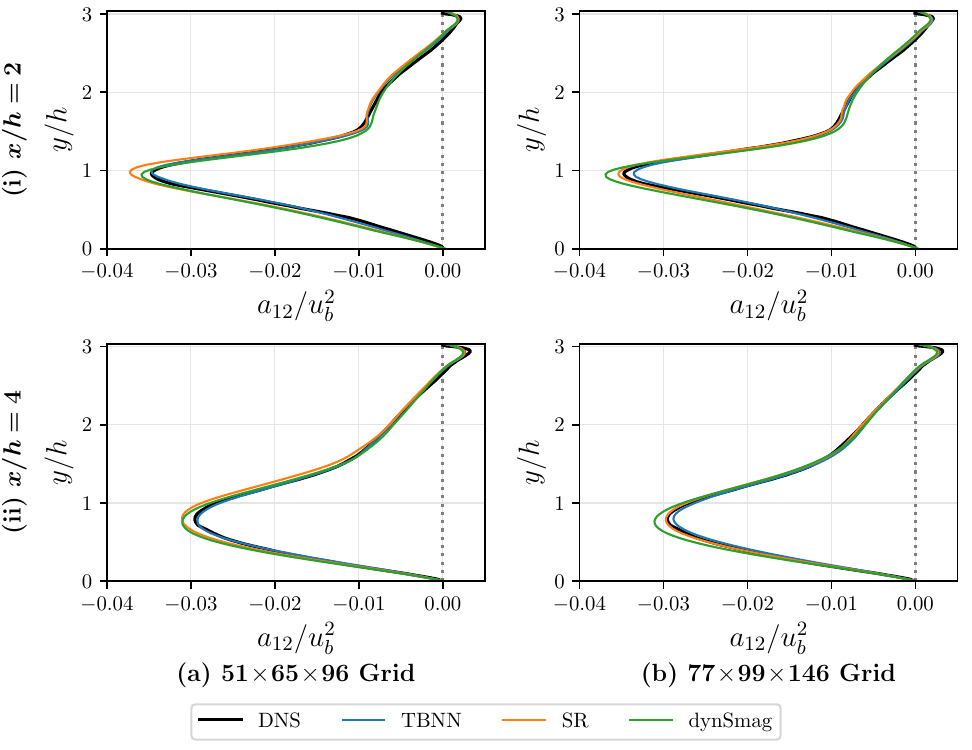}}% Images in 100% size
  \caption{Shear Reynolds stress component profiles for the periodic hill at $Re_h=5,600$ on (a) the coarser and (b) the finer grid resolutions at (\romannumeral1\relax) $x/h=2$ and (\romannumeral2\relax) $x/h=4$.}
    \label{fig:result - Hill RS uv}
\end{figure}

Finally, the shear Reynolds stress component, $a_{12}$, is shown in Figure~\ref{fig:result - Hill RS uv}. At both $x/h=2$ and $x/h=4$ on the coarser grid, the TBNN closely follows the DNS distribution, reproducing the near-wall slope, the profile peak, the transition through the bulk flow, and the trend near the top-wall. The other models maintain reasonably close predictions to the reference, but overpredict the peak magnitude. On the finer grid, the SR model provides the best agreement across these metrics, while the TBNN maintains a reasonable degree of accuracy, though a slightly damped prediction of the peak magnitude.

\end{document}